\documentclass{article}

    \PassOptionsToPackage{numbers, compress}{natbib}

\usepackage[dblblindworkshop,final]{neurips_2026}
\workshoptitle{Long-Context Foundation Models}

\usepackage[utf8]{inputenc} 
\usepackage[T1]{fontenc}    
\usepackage{hyperref}       
\usepackage{url}            
\usepackage{booktabs}       
\usepackage{amsfonts}       
\usepackage{nicefrac}       
\usepackage{microtype}      
\usepackage[table,xcdraw,dvipsnames]{xcolor}         
\usepackage{microtype}

\title{Where Should a Stuck Agent Spend Its Next Tokens?\\
Causal Attribution of Inference-Time Resources}

\author{
Saiyue Lyu$^{1*\dagger}$,
Jerrold Huang$^{2*}$,
Yunbei Zhang$^{3}$,
Janet Wang$^{3}$,\\
{\bf Zhenting Wang}$^{4}$,
{\bf Yingqiang Ge}$^{4}$,
{\bf Haibo Zhang}$^{5}$\\
${^1}$University of British Columbia, ${^2}$University of Toronto,
${^3}$Tulane University,\\
${^4}$Rutgers University,
${^5}$Kyushu Institute of Technology\\
}

\usepackage{graphicx} 
\usepackage{amsmath}
\usepackage{amssymb}

\usepackage{amsmath,amsfonts,bm}

\def\eqref#1{equation~\ref{#1}}

\def\1{\bm{1}}

\DeclareMathAlphabet{\mathsfit}{\encodingdefault}{\sfdefault}{m}{sl}
\SetMathAlphabet{\mathsfit}{bold}{\encodingdefault}{\sfdefault}{bx}{n}

\hypersetup{
 colorlinks=true,
 citecolor=cyan,
 linkcolor=BurntOrange,
 urlcolor=cyan}
\usepackage{float}
\usepackage{amsthm}
\newtheorem{proposition}{Proposition}

\newcommand{\ame}[1]{\mathrm{AME}_{#1}}

\usepackage{wrapfig}
\usepackage{tikz}
\newcommand*\circled[1]{\tikz[baseline=(char.base)]{
            \node[shape=circle,draw,inner sep=0.5pt] (char) {#1};}}

\usepackage{tikz,lipsum}
\usepackage[most]{tcolorbox}
\usepackage{pgfplots}
\pgfplotsset{compat=1.18}
\usepgfplotslibrary{groupplots}
\usepackage{lineno}

\begin{document}

\maketitle

\begingroup
\renewcommand{\thefootnote}{\fnsymbol{footnote}}
\footnotetext[1]{Equal contribution.}
\footnotetext[2]{Correspondence to: \texttt{saiyue.lyu@ubc.ca}}
\endgroup

\begin{abstract}
Inference-time scaling is often treated as a token-budget problem, but a failed coding agent can seek qualitatively different resources: computation, information, or feedback. We introduce Banzhaf Resource Attribution (BRA), a controlled factorial framework that measures each resource's marginal contribution to repair and its interactions while holding the failed state and revision protocol fixed. We evaluate six models on 194 function-level failures from MBPP+ and HumanEval+ and 126 repository-level failures from SWE-bench Verified. Task-relevant specifications consistently improve repair, whereas public-test feedback provides small gains and retrieved code little to none. The value of additional computation declines with model capability, while information remains useful; resource channels largely exhibit diminishing returns, although complementary information can reinforce itself. Yet three of four frontier models select feedback in $95$--$100\%$ of cases and obtain lower gains than uniform random resource selection. BRA thus exposes a gap between the resources agents prefer and those that measurably improve repair.
\end{abstract}


\section{Introduction}
\label{sec:intro}

Language models improve when given more to work with at inference time. Longer chains of thought~\citep{wei2022cot,openai2024o1,deepseek2025r1}, repeated sampling with a verifier~\citep{cobbe2021verifiers,lightman2024verify,brown2024monkeys}, and compute-optimal allocation between revision and search~\citep{snell2024scaling,wu2024inferencescaling} all trade tokens for accuracy along a single axis, the generation budget. For agents, however, inference-time resources differ in kind as well as in quantity. When a coding agent is stuck, its next tokens can buy three different things: more computation (deliberating without new observations), more information (documentation, retrieved code, a specification of what the program must do), or more feedback (executing the artifact and reading what failed)~\citep{yao2023react,lewis2020rag,chen2024selfdebug,shinn2023reflexion,yang2024sweagent}. Each channel has its own literature, evaluated by comparing end-to-end pipelines. Such comparisons do not isolate the contribution of the channel itself.


End-to-end comparisons cannot say which resource binds at a failed step, or whether a second still helps once the first is present, for three reasons. Along a trajectory the resource, the state, and the next revision vary together, so a verifier call that seems to help may only have triggered another edit. One-at-a-time ablations cannot tell sufficiency from necessity: rescuing a failure is not the same as being its bottleneck. And with bounded outcomes, an apparent interaction may be substitution, complementarity, or a ceiling unless every combination is observed.
 


These obstacles call for a controlled comparison of resources at fixed failed states. A resource-choice probe in coding tasks highlights why this comparison matters: when asked to choose a single resource before revising, three of the four frontier models we evaluate request execution feedback in $95$--$100\%$ of cases (\S\ref{sec:selfalloc}). 
Whether it is justified depends on how feedback compares with the other resources, alone and combined, at the same state. This leads to the central question of the paper: at a fixed failed state, how much does each inference-time resource contribute to a repair, how do the contributions combine, and how does the answer change with model capability?

To answer it we treat the three resources as players in a cooperative game whose worth is the utility an agent attains when a given bundle of them is supplied at the failed state. Asked this way, ``what is computation worth here?'' has no single answer: its marginal contribution depends on whether information and feedback are already present, and the natural summary is the average of that contribution over all bundles of the other resources, which is the resource's Banzhaf value~\citep{banzhaf1965}; the departures of the bundles' worths from additivity are the pairwise and three-way synergies of the game~\citep{harsanyi1963,grabisch1997}. Trajectories never reveal these quantities because an agent visits one bundle per state. 
We therefore introduce {\bf Banzhaf Resource Attribution (BRA)}: a framework for measuring how inference time resources contribute to repair at a fixed failed state. BRA first replicates this checkpoint across a full $2^n$ factorial design, with $n=3$ resource channels: computation ($C$), information ($I$), and feedback ($F$). 
Each replica receives exactly the resources of its cell and then one final revision call under a common protocol: the same model, prompt template, and output budget. Artifacts are scored on held-out tests disjoint from the public tests used for feedback. Because every bundle's worth is observed, Banzhaf values, interactions, and the Möbius decomposition of the full-resource gain are computed exactly per checkpoint and model, instead of estimated from samples.

We run BRA on 194 frozen function-level failures from MBPP+ and HumanEval+~\citep{austin2021mbpp,chen2021humaneval,liu2023evalplus} and 126 repository-level failures from SWE-bench Verified~\citep{jimenez2024swebench}, with six repair models spanning an order of magnitude of capability (Qwen3-4B/14B/32B, DeepSeek-V3.2, GPT-5.6, Claude Sonnet~4.5). Three findings recur. \circled{1} Information is the binding resource, and only the right information. An audited specification raises held-out utility for every model, by $0.06$--$0.14$ at function level and $0.05$--$0.10$ at repository level. A length-matched irrelevant packet is worth nothing, and so is retrieved repository code. Public-test feedback is worth at most $0.03$. \circled{2} More capable models gain less from computation, but still benefit from information. Along the ladder, the value of computation falls from $0.131$ to $0.007$; the value of information stays between $0.06$ and $0.14$. \circled{3} Resource channels show diminishing returns, while information components complement one another. All pairwise interactions among $C$, $I$, and $F$ are negative or null. Knowing where to change code is worth nothing alone; knowing what to change is worth about half the full effect; both together are worth the whole. These measurements also reveal a mismatch in resource selection: models exhibiting feedback fixation obtain lower average repair gains from their choices than from uniform random selection among the same resource options.

Our main contributions are: \circled{1} BRA, a controlled counterfactual design with pre-registered estimands, leakage audits, a placebo, and dose ladders, released with the frozen checkpoints and per-cell outcomes (\S\ref{sec:method}). \circled{2} Evidence across two task families and six models that inference resources are non-fungible and substitutable, that information rather than feedback binds, and that scaling migrates the bottleneck from computation to information (\S\ref{sec:results}). \circled{3} A decision-theoretic use of the surface that quantifies the misallocation of current agents. Using the same surfaces to score the agents' own resource choices, we find that three of four frontier models almost always request feedback, the least valuable resource, and realise less than a random choice would (\S\ref{sec:selfalloc}).

\begin{figure*}[t!]
    \centering
    \includegraphics[width=\linewidth]{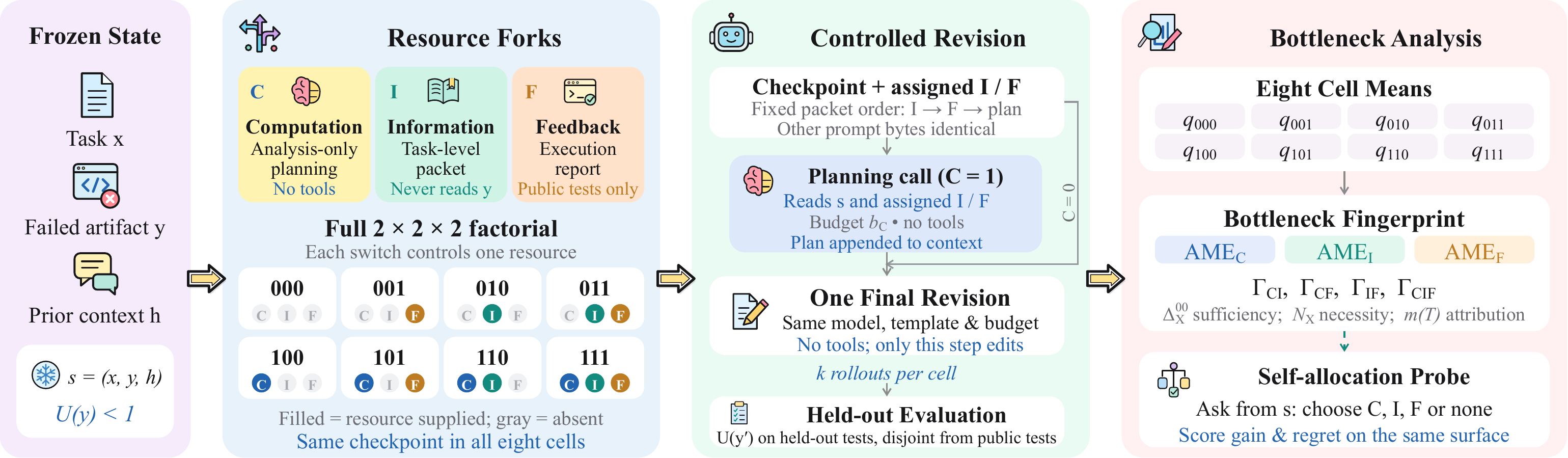}
    \caption{\textbf{BRA overview.}
 A frozen failed checkpoint $s=(x,y,h)$ is the unit of analysis. Computation, information, and feedback enter as separate interventions, each through its own channel, and every bundle of the three is assigned to the same $s$; a common final revision and held-out scoring identify the response surface, the eight cell means $q_{cif}$. A resource's value is its marginal contribution averaged over every bundle of the other two, its Banzhaf value; the interactions tell substitutes from complements. The fingerprint compares models on the same states and scores the agent's own resource choice by its regret.}
    \label{fig:overview}
\end{figure*}

\section{Preliminaries}
\label{sec:setup}
\vspace{-0.3cm}
A \emph{checkpoint} $s=(x,y,h)$ records a task, the artifact an agent has produced for it, and the context the agent saw; we consider checkpoints whose artifact fails the held-out evaluator, $U(y)<1$, and ask what would happen if the agent were granted a set of resources before revising once. Let $N=\{C,I,F\}$ be the three resources and, for a bundle $T\subseteq N$, let $Y_s(T)$ be the utility $s$ would attain if exactly the resources in $T$ were supplied and a fixed revision by a \emph{repair model} $M$; its expectation over the revision's randomness is $v_s(T)=\mathbb{E}[Y_s(T)]$. In a trajectory one of the eight potential outcomes is realised, chosen by the agent, so differences between trajectories are not causal effects.  
Because a checkpoint is frozen we can \emph{assign} every bundle to the same $s$ and observe all eight up to revision noise: a $2^3$ full factorial with the checkpoint as block~\citep{fisher1935design}, in which every interaction is estimable without assumptions.

{\bf The resource game.}
The expected outcomes $v_s(T)$ define a cooperative game on the player set $N$, in which a bundle's \emph{worth} is the utility obtained when exactly its resources are supplied; $g_s(T)=v_s(T)-v_s(\emptyset)$ is the corresponding \emph{gain game} with $g_s(\emptyset)=0$. The \emph{marginal contribution} of $X\notin T$ is $\mathrm{MV}_X(T)=v_s(T\cup\{X\})-v_s(T)$. Three classical objects summarise the game. The \emph{Banzhaf value}~\citep{banzhaf1965,dubey1979} and the \emph{Shapley value}~\citep{shapley1953} of resource $X$ are weighted averages of its marginal contributions over the bundles $T\subseteq N\setminus\{X\}$,
\begin{equation}
\beta_X(v_s)=\frac{1}{2^{n-1}}\sum_{T}\mathrm{MV}_X(T),\qquad
\phi_X(v_s)=\sum_{T}\frac{|T|!\,(n-1-|T|)!}{n!}\,\mathrm{MV}_X(T);
\label{eq:values}
\end{equation}
the Harsanyi dividends are the M\"obius coefficients $m(T)=\sum_{S\subseteq T}(-1)^{|T|-|S|}v_s(S)$, which attribute every unit of $g_s(N)=\sum_{T\neq\emptyset}m(T)$ exactly once to a single resource or to the synergy of a set~\citep{harsanyi1963,grabisch1997}; and the \emph{interaction index} of a pair is the average over bundles $T$ of the others of the second difference $\Delta_{XY}v_s(T)=v_s(T\cup\{X,Y\})-v_s(T\cup\{X\})-v_s(T\cup\{Y\})+v_s(T)$~\citep{grabisch1999interaction}. A positive pairwise dividend marks \emph{complements}, a negative one \emph{substitutes}; since a ceiling on $U$ also produces substitution, interactions are read against the distance of $v_s(\emptyset)$ from the ceiling. The pre-registered estimands are exactly these objects (\S\ref{sec:estimands}).
For readability we index a bundle by its indicator vector $(c,i,f)\in\{0,1\}^3$ and write $q_{cif}=v_s(T)$, so that $q_{000}=v_s(\emptyset)$, $q_{010}=v_s(\{I\})$, and $q_{111}=v_s(N)$; when the model matters we write $q_M(s;c,i,f)$. Tables and figures label the eight cells by $cif$. The two notations denote the same eight numbers: $v_s$ is used when a formula ranges over bundles, $q$ when it names a cell.

{\bf Costs and allocation.}
With a cost $\kappa_X>0$ per resource, additive over bundles, the \emph{return} of $X$ given $T$ is $\rho_X(T)=\mathrm{MV}_X(T)/\kappa_X$; an allocation policy chooses a bundle maximising $v_s(T)$ subject to $\kappa(T)\le B$, or one next resource $\pi(s)\in N\cup\{\varnothing\}$, whose realised gain $G(\pi)=\mathbb{E}_s[\mathrm{MV}_{\pi(s)}(\emptyset)]$ falls short of the oracle gain $G^\star=\mathbb{E}_s[\max_X\mathrm{MV}_X(\emptyset)]$ by the regret $R(\pi)$. $v_s$ is \emph{submodular} exactly when $m(\{X,Y\})\le0$ and $m(\{X,Y\})+m(N)\le0$ for every pair, in which case greedy allocation by return is provably near-optimal~\citep{nemhauser1978,khuller1999budgeted} (Appendix~\ref{app:game}).

\begin{tcolorbox}[colback=gray!10, colframe=gray!40,coltitle=black,title={\textbf{[Problem]} Inference-bottleneck attribution.}]
{\bf Given:} a population of frozen failed checkpoints, a held-out utility $U$, a set of repair models, and resource interventions with specified definitions and doses, with outcomes defined under a fixed final revision.
{\bf Estimate:} for each checkpoint $s$ and repair model $M$, the response surface $v^M_s(T)$ over bundles $T\subseteq\{C,I,F\}$, where $v^M_s(T)$ is the expected held-out utility after assigning $T$; from it, each resource's Banzhaf value and its interactions with the others (the fingerprint $z_s$). Across checkpoints, characterise their population averages and heterogeneity; across models, compare them on the same failed states.
{\bf Use:} evaluate a resource-allocation policy $\pi(s)$ by comparing the gain of its selected resource with that of a checkpoint-wise oracle over the same feasible choices, i.e.\ its allocation regret.
\end{tcolorbox}


\section{Banzhaf Resource Attribution (BRA)}
\label{sec:method}

Above problem asks for a surface that trajectories never reveal, because in a trajectory the resource, the state, and the revision move together. BRA makes it identifiable as illustrated in Figure~\ref{fig:overview}: clone the frozen state into every resource combination and follow each clone with one identical revision (\S\ref{sec:design}); define each resource so that it is supplied \emph{only} through its own channel, at a controlled dose (\S\ref{sec:treatments}); reduce the eight cell means to pre-registered contrasts with checkpoint-level uncertainty (\S\ref{sec:estimands}); and use the same surfaces to compare models and to score the agent's own allocation (\S\ref{sec:panels}).


\subsection{Checkpoint-conditioned counterfactual forks}
\label{sec:design}

A checkpoint is cloned into the eight cells of the full factorial over $(c,i,f)$; in each cell the model receives $s$ with exactly the indicated resources and then makes one common final revision, a single call with the same model, template, and maximum output length and no tools, whose output $y'$ replaces the artifact and is scored. Because the revision is identical across cells, any difference between cells is attributable to what preceded it, and because the intervention is assigned rather than chosen, $q_M(s;c,i,f)$ is identified by the cell mean, estimated from $k$ rollouts. Ablations from the empty cell identify only the simple effects $\Delta^{00}_X$; whether $X$ still has value once another resource is present is an interaction question, and the full factorial is the smallest design in which every interaction is estimable, at $8k$ revision and $4k$ planning calls per checkpoint and model (Appendix~\ref{app:cost}).

\subsection{Three resources, orthogonal as interventions and graded as doses}
\label{sec:treatments}

Computation, information, and feedback are not orthogonal in nature: a failing test is also information, and deliberation may reconstruct what a document would supply. They can be made orthogonal as interventions by fixing where each may come from. The definitions were frozen before any outcome was inspected and used unchanged for both task families.

\textbf{Computation} ($c=1$) is one analysis-only call by $M$ with a fixed output budget ($2048$ tokens in the main runs; $256$--$4096$ in the dose curve) that reads the task, the artifact, and the cell's other packets and returns a diagnosis and repair plan in prose, appended to the revision context; it may not execute, test, retrieve, or edit. All models run without hidden reasoning, so this call is the only extra computation in the design.

\textbf{Information} ($i=1$) is a packet that depends on the task but not on the artifact: generated once per task, identical across checkpoints and cells. Artifact independence is what separates information from feedback. Information is graded on a ladder of artifact-independent rungs: retrieval, deterministic BM25 retrieval over the repository at the base commit queried with the issue text; location, the files and enclosing functions the reference fix changes, extracted mechanically from the reference patch; behaviour, an account of the root cause and required behaviour written from the reference fix by a fixed writer model that may not name any file, function, or variable; and specification, both halves together (at function level, the input--output contract and edge cases). Written packets pass an automated audit against code, reference-solution lines, test literals and identifiers, and, for the behaviour rung, paths and symbols of the reference patch. A placebo arm supplies a length-matched neutral text, so the information effect reads as the value of content rather than presence; rungs above retrieval are instruments that establish what information would do, not what a system can obtain.

\textbf{Feedback} ($f=1$) is a report from executing the current artifact against a public evaluator, appended to the context; the artifact is never modified. Public tests are disjoint from the held-out tests defining $U$. Levels increase in detail: pass counts (F1), failing inputs (F2), with expected outputs (F3, the main level), with observed outputs (F4). On repositories, where public tests are the benchmark's own, a split-test protocol designates one of at least two failing target tests as observed, reports its outcome and failure excerpt, and scores $U$ on the rest; otherwise only outcome counts are reported.

Cells thus differ only in the presence of a packet; because $C$ may read the $I$ and $F$ packets, an interaction between computation and another resource measures whether deliberation makes that resource more productive; and $M$ is a technology, not a fourth resource, so the same intervention under two models yields two surfaces whose difference is their differential ability to convert resources into success.

\subsection{Estimands, fingerprint, and inference}
\label{sec:estimands}

Write $q_{cif}$ for $q_M(s;c,i,f)$ at a fixed checkpoint and model. The \emph{average main effect} of computation and the computation--information \emph{interaction}, each averaged over the remaining factors, are
\begin{equation}
\ame{C}(s)=\tfrac14\sum_{i,f}\big(q_{1if}-q_{0if}\big),\qquad
\Gamma_{CI}(s)=\tfrac12\sum_{f}\big(q_{11f}-q_{10f}-q_{01f}+q_{00f}\big),
\label{eq:ame}
\end{equation}
with $\ame{I}$, $\ame{F}$, $\Gamma_{CF}$, $\Gamma_{IF}$ analogous and $\Gamma_{CIF}=q_{111}-q_{110}-q_{101}-q_{011}+q_{100}+q_{010}+q_{001}-q_{000}$. The \emph{simple effect} $\Delta^{00}_X=\mathrm{MV}_X(\emptyset)$ measures standalone sufficiency and the \emph{conditional necessity} $N_X=\mathrm{MV}_X(N\setminus\{X\})$ what is lost when $X$ is removed from the full bundle; a resource sufficient alone but unnecessary given the others is a substitute for them. These estimands are the objects of the resource game of \S\ref{sec:setup}:
\begin{proposition}[Game-theoretic identity of the estimands]
\label{prop:game}
For every checkpoint and model, (i) $\ame{X}=\beta_X(v_s)$, the Banzhaf value of resource $X$; (ii) $\Gamma_{XY}$ equals the Banzhaf interaction index of $\{X,Y\}$, which for three resources coincides with the Shapley interaction index and equals $m(\{X,Y\})+\tfrac12 m(N)$; (iii) $\Gamma_{CIF}=m(N)$; (iv) $\sum_X\phi_X=q_{111}-q_{000}$ exactly, while $\sum_X\ame{X}=q_{111}-q_{000}-\tfrac14\Gamma_{CIF}$; and (v) the map from the eight cell means to $(q_{000},\ame{C},\ame{I},\ame{F},\Gamma_{CI},\Gamma_{CF},\Gamma_{IF},\Gamma_{CIF})$ is a linear bijection.
\end{proposition}

Proofs are short M\"obius algebra (Appendix~\ref{app:game}). We call the seven contrasts $z_s=[\ame{C},\ame{I},\ame{F},\Gamma_{CI},\Gamma_{CF},\Gamma_{IF},\Gamma_{CIF}](s)$ the checkpoint's \emph{bottleneck fingerprint}. By (v), $z_s$ together with $q_{000}$ carries the same information as the eight cell means. By (iv), the three main effects account for the whole gain $q_{111}-q_{000}$ except a quarter of the three-way term, which is small in practice; the Shapley values account for it exactly and rank the resources identically on every model (Table~\ref{tab:shapley}). Each checkpoint is also labelled by a rule fixed in advance: \emph{$X$-responsive} if resource $X$ alone lifts $U$ by at least $\tau=0.2$ over the empty cell; \emph{jointly unlocked} if no single resource does but some combination does; \emph{unresolved} if no cell does; and \emph{degenerate} if the empty cell is already near the ceiling ($q_{000}\ge0.8$) (Appendix~\ref{app:tables}).
The unit of replication is the checkpoint. Every contrast is computed per checkpoint from the $k$-rollout cell means, averaged across checkpoints, and given a confidence interval by resampling checkpoints~\citep{cameron2008bootstrap}; the three pairwise interactions are Holm-corrected~\citep{holm1979}. The primary outcome is the held-out pass fraction $U$; the binary ``all held-out tests pass'' is secondary. Dose comparisons are paired by checkpoint.

\subsection{Comparing models and scoring the agent's own allocation}
\label{sec:panels}

Two panels separate two readings of capability. In the fixed-state panel all models repair the same checkpoints, so differences lie in $q_M$ alone; a matched-difficulty subset keeps only checkpoints every model fails badly. In the native-failure panel each model repairs its own first-attempt failures, so the population shifts with model $M$. A within-family size ladder checks the vendor confound.

{\bf Self-allocation probe.}
\label{sec:probe}
Given the empty cell's context, the model chooses one of $C$ (a planning pass), $I$ (a specification of the required behaviour), $F$ (the execution report on its artifact), or nothing. Its choices form a policy $\pi$, scored on the same surfaces by $G(\pi)$, $G^\star$, and $R(\pi)$ of \S\ref{sec:setup} against ``always $X$'' and a uniformly random choice; option order is permuted to separate preference from position.

{\bf Validity controls.}
Only the final revision may change the artifact; the computation call has no tools; the information packet never reads the artifact; the feedback step never searches. Held-out inputs, expected outputs, and reference solutions never enter a prompt, which a canary test asserts per cell. The final call's template and limit are identical across cells, and truncated revisions are scored, not retried. Definitions, contrasts, the label rule, and manifests are frozen and hashed before any outcome is read; later changes are logged as deviations (Appendix~\ref{app:deviations}). All cells of an experiment are scored by one sandboxed evaluator on one machine.


\section{Experimental Setup}
\label{sec:experiments}

We run BRA on two task families and six models to answer: \textbf{Q1} which resource binds at a failed step; \textbf{Q2} how the fingerprint changes with capability; \textbf{Q3} whether resources complement or substitute; \textbf{Q4} whether the picture transfers to repositories, and which \emph{kind} of information binds there; \textbf{Q5} whether models request the resource that would help them; \textbf{Q6} whether tokens are a common currency and greedy allocation is near-optimal.

{\bf Function-level checkpoints.}
Every failed first attempt of Qwen3-4B (quick-draft style, $2048$ tokens, temperature $0.7$) on MBPP+ and HumanEval+~\citep{austin2021mbpp,chen2021humaneval,liu2023evalplus}: $N=194$ ($131$ MBPP+, $63$ HumanEval+; mean $U=0.32$). $U$ is the fraction of EvalPlus's extended tests passed; public tests are the datasets' original assertions, disjoint from the held-out set.

{\bf Repository-level checkpoints.}
On a stratified frame of $250$ SWE-bench Verified instances~\citep{jimenez2024swebench} a text-based mini-SWE-agent with Qwen3-14B (75 steps) submitted $126$ unresolved patches; these are the checkpoints ($62$ django, $21$ sympy, $43$ from eight other repositories; in $57$ the patch breaks the test run, in $69$ target tests fail). $U$ is the fraction of held-out FAIL\_TO\_PASS tests passed, $0$ on any PASS\_TO\_PASS regression; $33$ instances with two or more target tests use the split-test protocol. A native-failure panel uses $30$ further checkpoints on which DeepSeek-V3.2, run as the agent, failed.

{\bf Models and treatments.}
Qwen3-4B, 14B, and 32B form a within-family ladder; DeepSeek-V3.2, GPT-5.6, and Claude Sonnet~4.5 are frontier models. All run without hidden reasoning, at temperature $0.7$ (nucleus $0.8$ where supported). Qwen3-14B and DeepSeek-V3.2 use $k=5$ rollouts per cell at function level, all other runs $k=3$. Computation is a $2048$-token planning call (dose $256$--$4096$); feedback is F3 with at most five failing cases (dose F1--F4); function-level information is the specification rung written by Kimi~K2.5 (never a repair model), with a length-matched placebo; repository-level information is the retrieval rung in the main runs and the location, behaviour, and specification rungs in dose arms that re-run the four $I{=}1$ cells. The final revision has $2048$ output tokens at function level and $4096$ at repository level, where it is expressed as search/replace edits scored with the official harness (Appendix~\ref{app:swe}). Intervals are checkpoint bootstraps with $2000$ resamples; Holm correction applies within the three pairwise interactions; labels use $\tau=0.2$ and ceiling $0.8$. Compute is reported in Appendix~\ref{app:cost}.


\section{Results}
\label{sec:results}

\begin{figure}[t]
\centering
\includegraphics[width=\linewidth]{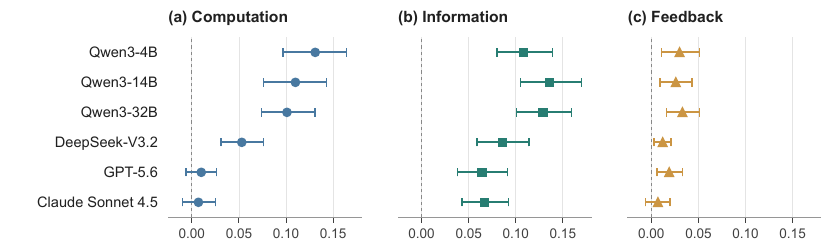}
\vspace{-0.5cm}
\caption{\textbf{Resource effects.} Average main effects on held-out utility across 194 function-level checkpoints. Bars show 95\% checkpoint-bootstrap intervals. Dashed lines mark zero.}
\label{fig:resource_effects}
\end{figure}
\vspace{-0.2cm}

\subsection{Q1: Information is the binding resource; feedback is not}
\label{sec:q1}

Figure~\ref{fig:resource_effects} gives the population estimates for six models on the same 194 checkpoints. Information has the largest main effect for five of the six models, $0.064$ (GPT-5.6) to $0.136$ (Qwen3-14B), at $p<0.001$ throughout; feedback from public tests (F3) is worth $0.007$--$0.033$; computation lies between the two for weak models and vanishes for strong ones (\S\ref{sec:q2}). The information effect is content, not length: a length-matched neutral packet changes utility by $-0.016$ (Qwen3-14B) and $-0.002$ (DeepSeek-V3.2), neither distinguishable from zero, while the real packet exceeds it by $0.195$ and $0.126$ (Table~\ref{tab:placebo}). Nor is feedback under-dosed: on the 40 hardest checkpoints (DeepSeek-V3.2, empty cell $0.298$; Figure~\ref{fig:dose}) the feedback-only cell rises only from $0.362$ (pass counts, F1) to $0.402$ (full observed outputs, F4), whereas even a $256$-token plan reaches $0.489$.

\noindent\textbf{Takeaway (Q1).} \emph{At a failed step the binding resource is information about what the code must do. Public-test feedback is worth at most $0.03$ and stays small at its most detailed level; the information effect is content, since a length-matched placebo is worth nothing.}

\subsection{Q2: Capability moves the bottleneck from computation to information}
\label{sec:q2}

Reading down Figure~\ref{fig:resource_effects}, ordered by baseline capability, the computation effect falls monotonically: $0.131$, $0.110$, $0.101$ along the Qwen3 ladder, $0.053$ for DeepSeek-V3.2, then $0.010$ and $0.007$ for GPT-5.6 and Sonnet~4.5, whose intervals include zero. The information effect falls far less, from at most $0.136$ to $0.064$, and remains the largest for five of the six models; its share of the summed main effects rises from $40\%$ to $83\%$ while computation's falls from $49\%$ to $8\%$ (Table~\ref{tab:function_main}).

Two objections must be met. First, a ceiling could dilute the strong models' effects, since they solve more checkpoints outright. On the 24 checkpoints every model fails badly ($q_{000}<0.5$ for all six), information is worth $0.12$--$0.25$ and is significant for every model, so its decline in the full sample is a ceiling effect; computation is not rescued, at $0.03$ and not significant for GPT-5.6 and Sonnet against $0.06$--$0.10$ for the other four models (Figure~\ref{fig:ceiling}, top). Second, capability is confounded with vendor; the within-family ladder Qwen3-4B $\to$ 14B $\to$ 32B shows the same direction (computation $0.131\to0.101$, information $0.11$--$0.14$), though the largest step occurs between 32B and the frontier models.

\noindent\textbf{Takeaway (Q2).} \emph{Scaling moves inference bottlenecks rather than removing them. Extra computation is worth $0.13$ to the weakest model and nothing to the strongest, on the same and on matched-difficulty states, while information keeps its value on matched states; for the strongest models information is by far the largest remaining lever.}

\subsection{Q3: Resources substitute for one another}
\label{sec:q3}

The design was pre-registered with a complementarity hypothesis, that computation would raise the marginal value of information and feedback; the data reject it. Across six models, three pairs, and two samples (all checkpoints, and those with $q_{000}<0.5$), $21$ of the $36$ pairwise interactions are significantly negative after Holm correction and none is significantly positive (Table~\ref{tab:function_interactions}). On all checkpoints the clearest case is $\Gamma_{CF}$ along the Qwen ladder ($-0.090$, $-0.056$, $-0.045$): a plan and a test report recover the same failures. Away from the ceiling substitution becomes \emph{stronger}, not weaker: $\Gamma_{CI}$ lies between $-0.10$ and $-0.14$ and is significant for every model but Qwen3-4B (Figure~\ref{fig:ceiling}, bottom), so the negative terms are not an artifact of utility saturating near $1$.

\noindent\textbf{Takeaway (Q3).} \emph{Computation, information, and feedback are substitutes, more strongly so away from the ceiling: a second resource adds less than it would alone, so the first choice matters most.}

\subsection{Q4: On repositories, retrieval is not information}
\label{sec:q4}

\begin{figure}[t]
\centering
\includegraphics[width=\linewidth]{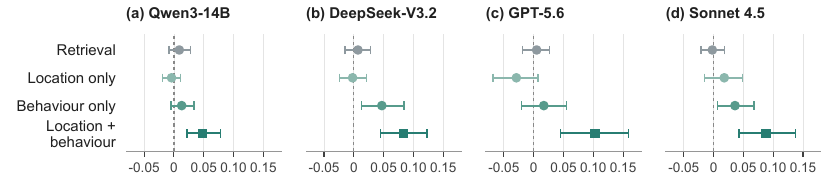}
\vspace{-0.5cm}
\caption{\textbf{Information content.}
Points show average information effects on held-out utility
across 126 repository-level checkpoints ($k=3$);
bars show 95\% checkpoint-bootstrap intervals.}
\label{fig:information_types}
\end{figure}

\begin{table*}[t]
\centering
\caption{Repository-level fixed-state panel ($N=126$, $k=3$): empty and full-resource cells and average main effects with 95\% intervals. Left: BM25-retrieved repository context as the information packet. Right: the same $I{=}0$ cells with the $I{=}1$ cells re-run under an audited oracle specification of the fix. Bold: $p<0.05$. Interactions, artifact strata, and the native-failure panel are extended in Table~\ref{tab:swe_robust}.}
\label{tab:swe_main}
\resizebox{\linewidth}{!}{\begin{tabular}{l r r r r r r r r r}
\toprule
& & \multicolumn{4}{c}{retrieval information} & \multicolumn{4}{c}{audited specification}\\
\cmidrule(lr){3-6}\cmidrule(lr){7-10}
Model & $q_{000}$ & $q_{111}$ & $\mathrm{AME}_C$ & $\mathrm{AME}_I$ & $\mathrm{AME}_F$ & $q_{111}$ & $\mathrm{AME}_C$ & $\mathrm{AME}_I$ & $\mathrm{AME}_F$\\
\midrule
Qwen3-14B & 0.026 & 0.059 & 0.016 {\scriptsize[$-$0.005, 0.039]} & 0.009 {\scriptsize[$-$0.008, 0.028]} & 0.001 {\scriptsize[$-$0.020, 0.020]} & 0.125 & \textbf{0.034} {\scriptsize[0.012, 0.057]} & \textbf{0.048} {\scriptsize[0.022, 0.078]} & 0.008 {\scriptsize[$-$0.011, 0.026]}\\
DeepSeek-V3.2 & 0.238 & 0.275 & 0.039 {\scriptsize[$-$0.001, 0.078]} & 0.007 {\scriptsize[$-$0.015, 0.028]} & $-$0.006 {\scriptsize[$-$0.027, 0.014]} & 0.359 & \textbf{0.034} {\scriptsize[0.006, 0.064]} & \textbf{0.083} {\scriptsize[0.045, 0.123]} & 0.011 {\scriptsize[$-$0.012, 0.035]}\\
GPT-5.6 & 0.423 & 0.447 & 0.024 {\scriptsize[$-$0.005, 0.054]} & 0.005 {\scriptsize[$-$0.019, 0.027]} & $-$0.004 {\scriptsize[$-$0.027, 0.020]} & 0.538 & 0.009 {\scriptsize[$-$0.016, 0.037]} & \textbf{0.103} {\scriptsize[0.045, 0.159]} & 0.009 {\scriptsize[$-$0.010, 0.028]}\\
Claude Sonnet 4.5 & 0.324 & 0.338 & 0.014 {\scriptsize[$-$0.016, 0.047]} & $-$0.002 {\scriptsize[$-$0.021, 0.018]} & 0.009 {\scriptsize[$-$0.012, 0.030]} & 0.392 & $-$0.024 {\scriptsize[$-$0.053, 0.004]} & \textbf{0.088} {\scriptsize[0.043, 0.137]} & 0.001 {\scriptsize[$-$0.017, 0.019]}\\
\bottomrule
\end{tabular}
}
\vspace{-2mm}
\end{table*}

Repository-level failures are much harder for one revision (empty cell $0.03$--$0.42$). With the information a deployed agent actually receives, five retrieved code chunks, \emph{no resource has a significant main effect for any model} (Table~\ref{tab:swe_main}, left): retrieval is worth $-0.002$ to $0.009$ and feedback $-0.006$ to $0.009$. Replacing the retrieved chunks by an audited specification of the fix changes the picture entirely (right): the information effect becomes $0.048$, $0.083$, $0.088$, and $0.103$ for Qwen3-14B, DeepSeek-V3.2, Sonnet, and GPT-5.6, all at $p\le0.002$, and the full-resource cell exceeds the empty cell by $0.07$--$0.12$ instead of $0.01$--$0.04$; computation stays small and confined to the weaker models, feedback nil. The pattern holds in both artifact strata, where the specification effect is positive for every model and retrieval is at most $0.033$, and on 30 checkpoints DeepSeek-V3.2 itself had failed as the agent (specification $0.079$, retrieval $-0.004$); interactions are small throughout (Table~\ref{tab:swe_robust}).

The dose ladder locates what retrieval misses (Figure~\ref{fig:information_types}). Knowing \emph{where} the fix goes is worth nothing alone ($-0.03$ to $+0.02$), knowing \emph{what} it must do is worth $0.013$--$0.047$, or $17$--$57\%$ of the full effect, and both together the whole ($0.048$--$0.103$). The two halves complement: $\ame{I}^{\text{spec}}-\ame{I}^{\text{loc}}-\ame{I}^{\text{beh}}$ is positive for all four models ($0.034$--$0.115$), with a bootstrap interval excluding zero for three (Sonnet~4.5: $p=0.09$).

\noindent\textbf{Takeaway (Q4).} \emph{On repository-level failures the information a deployed agent gets, retrieved code, is worth nothing, and so is feedback; the same states become tractable when the model is told what the code must do and where, and the two halves are worth more together than apart. Information binds on repositories too, but retrieval does not supply it.}

\subsection{Q5: Models ask for the wrong resource}
\label{sec:selfalloc}

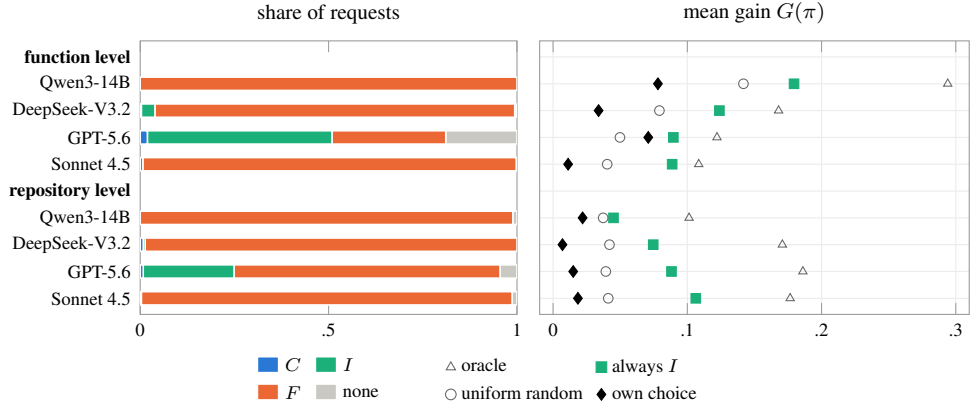
\begin{figure}[t]
\centering
\definecolor{cC}{HTML}{2A78D6}
\definecolor{cI}{HTML}{1BAF7A}
\definecolor{cF}{HTML}{EB6834}
\definecolor{cN}{HTML}{8A8984}
\definecolor{cG}{HTML}{C9C8C3}
\begin{tikzpicture}
\begin{groupplot}[
  group style={group size=2 by 1, horizontal sep=3mm, yticklabels at=edge left},
  height=52mm, every axis/.append style={font=\footnotesize}, y dir=reverse, ymin=-0.6, ymax=9.6,
  ytick={0,1,2,3,4,5,6,7,8,9}, yticklabels={\textbf{function level},Qwen3-14B,DeepSeek-V3.2,GPT-5.6,Sonnet 4.5,\textbf{repository level},Qwen3-14B,DeepSeek-V3.2,GPT-5.6,Sonnet 4.5}, tick label style={font=\scriptsize}, y tick style={draw=none},
  axis line style={black!40}, tick style={black!40}, title style={font=\footnotesize, yshift=-1mm},
  legend style={draw=none, font=\scriptsize, legend columns=2, /tikz/every even column/.append style={column sep=3pt}},
  legend cell align=left,
]
\nextgroupplot[xbar stacked, bar width=5pt, width=0.47\linewidth, xmin=0, xmax=1, xtick={0,0.5,1}, xticklabels={0,.5,1}, title={share of requests}, legend columns=2, legend to name=saleg1]
\addplot[fill=cC, draw=white, line width=0.6pt] coordinates {(0.000,1) (0.003,2) (0.019,3) (0.000,4) (0.000,6) (0.008,7) (0.008,8) (0.000,9)};
\addlegendentry{$C$}
\addplot[fill=cI, draw=white, line width=0.6pt] coordinates {(0.000,1) (0.036,2) (0.490,3) (0.007,4) (0.000,6) (0.005,7) (0.241,8) (0.003,9)};
\addlegendentry{$I$}
\addplot[fill=cF, draw=white, line width=0.6pt] coordinates {(1.000,1) (0.955,2) (0.302,3) (0.991,4) (0.989,6) (0.987,7) (0.706,8) (0.984,9)};
\addlegendentry{$F$}
\addplot[fill=cG, draw=white, line width=0.6pt] coordinates {(0.000,1) (0.005,2) (0.189,3) (0.002,4) (0.011,6) (0.000,7) (0.045,8) (0.013,9)};
\addlegendentry{none}
\nextgroupplot[width=0.52\linewidth, xmin=-0.01, xmax=0.31, xtick={0,0.1,0.2,0.3}, xticklabels={0,.1,.2,.3}, grid=major, grid style={black!8}, title={mean gain $G(\pi)$}, mark size=1.8pt, legend to name=saleg2]
\addplot[only marks, mark=triangle, color=black!60] coordinates {(0.2940,1) (0.1680,2) (0.1221,3) (0.1085,4) (0.1013,6) (0.1708,7) (0.1861,8) (0.1766,9)};
\addlegendentry{oracle}
\addplot[only marks, mark=square*, color=cI] coordinates {(0.1795,1) (0.1239,2) (0.0896,3) (0.0887,4) (0.0450,6) (0.0746,7) (0.0882,8) (0.1063,9)};
\addlegendentry{always $I$}
\addplot[only marks, mark=o, color=black!60, mark options={fill=white}] coordinates {(0.1418,1) (0.0792,2) (0.0498,3) (0.0403,4) (0.0373,6) (0.0420,7) (0.0393,8) (0.0411,9)};
\addlegendentry{uniform random}
\addplot[only marks, mark=diamond*, color=black, mark size=2.2pt] coordinates {(0.0781,1) (0.0339,2) (0.0709,3) (0.0112,4) (0.0220,6) (0.0070,7) (0.0150,8) (0.0185,9)};
\addlegendentry{own choice}
\end{groupplot}
\end{tikzpicture}

\pgfplotslegendfromname{saleg1}\hspace{6mm}\pgfplotslegendfromname{saleg2}
\vspace{-2mm}
\caption{\textbf{Self-allocation probe} ($3$ rollouts/checkpoint; repository level uses the specification rung for $I$). Left: share of requests for each resource. Right: mean realised gain $G(\pi)$ of the model's own choices, of always requesting information, of a uniformly random choice among $C$, $I$, $F$, and of the per-checkpoint oracle $G^\star$.}
\label{fig:selfalloc}
\end{figure}

Asked which single resource they want, Qwen3-14B, DeepSeek-V3.2, and Sonnet~4.5 request execution feedback in $95$--$100\%$ of checkpoints on both task families (Figure~\ref{fig:selfalloc}); GPT-5.6 asks for information $49\%$ of the time at function level but prefers feedback on repositories ($71\%$). Scored against each checkpoint's own surface, feedback fixation is expensive. At function level the three fixated models realise only $10$--$27\%$ of the oracle gain $G^\star$, fall short of always requesting information by $0.078$--$0.101$, and are significantly worse than a uniformly random choice (by $0.029$--$0.064$, $p<0.001$); GPT-5.6 is the only model not worse than random ($0.071$ against $0.050$, difference not significant). On repositories always-information beats the model's own choice for all four models, significantly for three, and every model falls below random in point estimate (significantly for DeepSeek-V3.2). Even always-information leaves gain on the table, because for $28$--$45\%$ of checkpoints the best single resource is not the population's best (Table~\ref{tab:function_main}).

\noindent\textbf{Takeaway (Q5).} \emph{
Three of the four probed models almost always request feedback, the least valuable resource, and forgo most of the attainable gain: at function level they do significantly worse than choosing at random, and the fixed rule "always request information" beats them on both tasks.
}

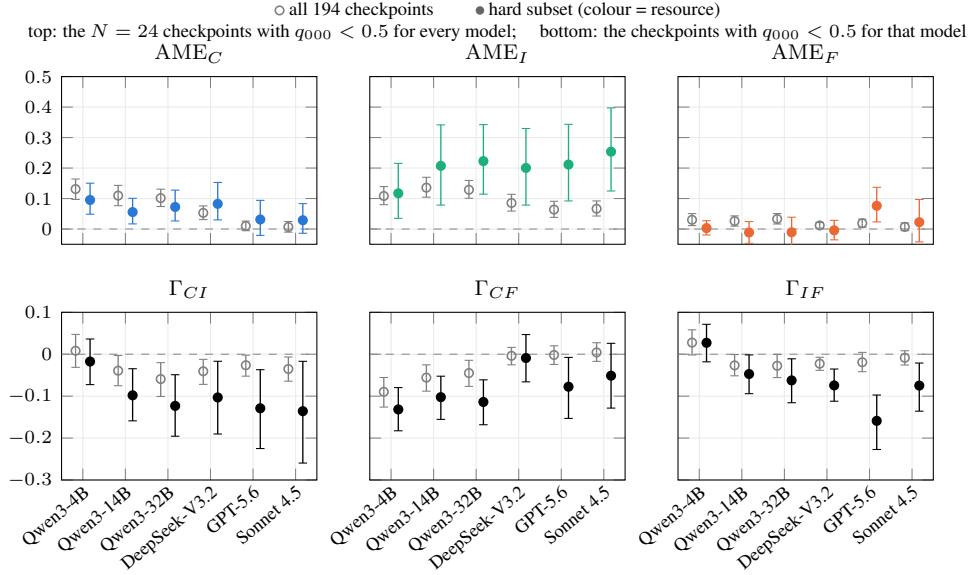
\begin{figure*}[!t]
\centering
\definecolor{cC}{HTML}{2A78D6}
\definecolor{cI}{HTML}{1BAF7A}
\definecolor{cF}{HTML}{EB6834}
\definecolor{cN}{HTML}{8A8984}
\definecolor{cG}{HTML}{C9C8C3}
\begin{tikzpicture}
\begin{groupplot}[
  group style={group size=3 by 2, horizontal sep=7mm, vertical sep=9mm, yticklabels at=edge left},
  width=0.355\linewidth, height=38mm, every axis/.append style={font=\footnotesize},
  xmin=0.5, xmax=6.5, xtick={1,...,6}, x tick label style={rotate=40, anchor=north east, inner sep=1pt, font=\scriptsize},
  ytick distance=0.1, y tick label style={font=\scriptsize, /pgf/number format/fixed}, grid=major, grid style={black!8},
  title style={font=\footnotesize, yshift=-1.5mm}, every axis plot/.append style={mark size=1.6pt, line width=0.6pt},
  error bars/y dir=both, error bars/y explicit, error bars/error bar style={line width=0.5pt},
]
\nextgroupplot[title={$\mathrm{AME}_C$}, ymin=-0.05, ymax=0.5, xticklabels={}]
\draw[black!45, dashed] (axis cs:0.5,0) -- (axis cs:6.5,0);
\addplot[color=black!50, mark=o, only marks, mark options={fill=white}] table[x=x, y=y, y error minus=lo, y error plus=hi] {
    x y lo hi
    0.83 0.1311 0.0342 0.0329
    1.83 0.1095 0.0330 0.0337
    2.83 0.1014 0.0276 0.0293
    3.83 0.0530 0.0220 0.0228
    4.83 0.0100 0.0155 0.0160
    5.83 0.0068 0.0170 0.0179
};
\addplot[color=cC, mark=*, only marks] table[x=x, y=y, y error minus=lo, y error plus=hi] {
    x y lo hi
    1.17 0.0953 0.0467 0.0551
    2.17 0.0556 0.0388 0.0451
    3.17 0.0724 0.0459 0.0552
    4.17 0.0826 0.0526 0.0702
    5.17 0.0314 0.0522 0.0626
    6.17 0.0288 0.0427 0.0545
};
\nextgroupplot[title={$\mathrm{AME}_I$}, ymin=-0.05, ymax=0.5, xticklabels={}]
\draw[black!45, dashed] (axis cs:0.5,0) -- (axis cs:6.5,0);
\addplot[color=black!50, mark=o, only marks, mark options={fill=white}] table[x=x, y=y, y error minus=lo, y error plus=hi] {
    x y lo hi
    0.83 0.1084 0.0286 0.0309
    1.83 0.1357 0.0312 0.0340
    2.83 0.1288 0.0279 0.0306
    3.83 0.0855 0.0266 0.0285
    4.83 0.0638 0.0255 0.0272
    5.83 0.0666 0.0240 0.0256
};
\addplot[color=cI, mark=*, only marks] table[x=x, y=y, y error minus=lo, y error plus=hi] {
    x y lo hi
    1.17 0.1171 0.0821 0.0982
    2.17 0.2072 0.1287 0.1344
    3.17 0.2227 0.1084 0.1199
    4.17 0.2004 0.1219 0.1294
    5.17 0.2115 0.1194 0.1319
    6.17 0.2537 0.1290 0.1436
};
\nextgroupplot[title={$\mathrm{AME}_F$}, ymin=-0.05, ymax=0.5, xticklabels={}]
\draw[black!45, dashed] (axis cs:0.5,0) -- (axis cs:6.5,0);
\addplot[color=black!50, mark=o, only marks, mark options={fill=white}] table[x=x, y=y, y error minus=lo, y error plus=hi] {
    x y lo hi
    0.83 0.0302 0.0190 0.0204
    1.83 0.0261 0.0166 0.0170
    2.83 0.0331 0.0166 0.0176
    3.83 0.0118 0.0089 0.0094
    4.83 0.0189 0.0130 0.0137
    5.83 0.0071 0.0126 0.0131
};
\addplot[color=cF, mark=*, only marks] table[x=x, y=y, y error minus=lo, y error plus=hi] {
    x y lo hi
    1.17 0.0031 0.0226 0.0241
    2.17 -0.0114 0.0354 0.0359
    3.17 -0.0108 0.0489 0.0496
    4.17 -0.0042 0.0313 0.0327
    5.17 0.0762 0.0529 0.0607
    6.17 0.0224 0.0646 0.0746
};
\nextgroupplot[title={$\Gamma_{CI}$}, ymin=-0.3, ymax=0.1, xticklabels={Qwen3-4B,Qwen3-14B,Qwen3-32B,DeepSeek-V3.2,GPT-5.6,Sonnet 4.5}]
\draw[black!45, dashed] (axis cs:0.5,0) -- (axis cs:6.5,0);
\addplot[color=black!50, mark=o, only marks, mark options={fill=white}] table[x=x, y=y, y error minus=lo, y error plus=hi] {
    x y lo hi
    0.83 0.0082 0.0395 0.0391
    1.83 -0.0392 0.0359 0.0362
    2.83 -0.0592 0.0416 0.0391
    3.83 -0.0407 0.0312 0.0285
    4.83 -0.0262 0.0261 0.0241
    5.83 -0.0352 0.0291 0.0286
};
\addplot[color=black, mark=*, only marks] table[x=x, y=y, y error minus=lo, y error plus=hi] {
    x y lo hi
    1.17 -0.0174 0.0551 0.0537
    2.17 -0.0980 0.0609 0.0634
    3.17 -0.1232 0.0724 0.0743
    4.17 -0.1031 0.0872 0.0863
    5.17 -0.1289 0.0961 0.0920
    6.17 -0.1358 0.1238 0.1189
};
\nextgroupplot[title={$\Gamma_{CF}$}, ymin=-0.3, ymax=0.1, xticklabels={Qwen3-4B,Qwen3-14B,Qwen3-32B,DeepSeek-V3.2,GPT-5.6,Sonnet 4.5}]
\draw[black!45, dashed] (axis cs:0.5,0) -- (axis cs:6.5,0);
\addplot[color=black!50, mark=o, only marks, mark options={fill=white}] table[x=x, y=y, y error minus=lo, y error plus=hi] {
    x y lo hi
    0.83 -0.0897 0.0364 0.0342
    1.83 -0.0558 0.0323 0.0305
    2.83 -0.0449 0.0318 0.0304
    3.83 -0.0038 0.0215 0.0201
    4.83 -0.0018 0.0225 0.0219
    5.83 0.0043 0.0213 0.0231
};
\addplot[color=black, mark=*, only marks] table[x=x, y=y, y error minus=lo, y error plus=hi] {
    x y lo hi
    1.17 -0.1317 0.0509 0.0521
    2.17 -0.1024 0.0530 0.0500
    3.17 -0.1139 0.0543 0.0528
    4.17 -0.0090 0.0569 0.0559
    5.17 -0.0775 0.0756 0.0699
    6.17 -0.0512 0.0773 0.0772
};
\nextgroupplot[title={$\Gamma_{IF}$}, ymin=-0.3, ymax=0.1, xticklabels={Qwen3-4B,Qwen3-14B,Qwen3-32B,DeepSeek-V3.2,GPT-5.6,Sonnet 4.5}]
\draw[black!45, dashed] (axis cs:0.5,0) -- (axis cs:6.5,0);
\addplot[color=black!50, mark=o, only marks, mark options={fill=white}] table[x=x, y=y, y error minus=lo, y error plus=hi] {
    x y lo hi
    0.83 0.0278 0.0295 0.0304
    1.83 -0.0266 0.0249 0.0259
    2.83 -0.0276 0.0281 0.0272
    3.83 -0.0226 0.0158 0.0154
    4.83 -0.0191 0.0226 0.0235
    5.83 -0.0085 0.0173 0.0168
};
\addplot[color=black, mark=*, only marks] table[x=x, y=y, y error minus=lo, y error plus=hi] {
    x y lo hi
    1.17 0.0273 0.0454 0.0441
    2.17 -0.0472 0.0469 0.0456
    3.17 -0.0624 0.0533 0.0515
    4.17 -0.0744 0.0377 0.0391
    5.17 -0.1589 0.0683 0.0616
    6.17 -0.0748 0.0610 0.0538
};
\end{groupplot}
\node[anchor=south, font=\scriptsize, align=center] at ($(group c2r1.north)+(0,3.5mm)$) {\tikz[baseline=-0.6ex]\draw[black!50, fill=white, line width=0.6pt] (0,0) circle (1.6pt); all 194 checkpoints \qquad \tikz[baseline=-0.6ex]\fill[black!60] (0,0) circle (1.6pt); hard subset (colour = resource)\\ top: the $N=24$ checkpoints with $q_{000}<0.5$ for every model; \quad bottom: the checkpoints with $q_{000}<0.5$ for that model};
\end{tikzpicture}
\vspace{-2mm}
\caption{Ceiling checks, function level, with 95\% checkpoint-bootstrap intervals. Top: main effects on all checkpoints (open) and on the $24$ checkpoints on which every model has $q_{000}<0.5$ (filled). Bottom: pairwise interactions on all checkpoints (open) and on the checkpoints with $q_{000}<0.5$ for that model (filled; $N$ in Table~\ref{tab:function_interactions}). Information stays largest away from the ceiling, and the interactions remain negative or near zero; because both subsets are selected on $q_{000}$, the size of the shifts is not an unbiased estimate (Appendix~\ref{app:tables}).}
\label{fig:ceiling}
\end{figure*}

\subsection{Q6: Return per cost, non-fungibility, and greedy allocation}
\label{sec:econ}

Are tokens a common currency? Within checkpoints, the resource indicators predict the eight cell means better than the cell's total token cost for every model and both task families, except GPT-5.6 at function level, where the two tie (five-fold cross-validated $R^2$; Table~\ref{tab:econ}); on repositories token cost alone predicts worse than the checkpoint mean. Per $1000$ added tokens, information returns $0.54$--$1.25$ at function level, four to over twenty times the return of computation, and $0.11$--$0.31$ on repositories; feedback is cheap, so its per-token return is respectable despite its small gain. Allocating greedily by return reaches $0.80$--$0.99$ of the optimal affordable bundle's gain at function level and $0.89$--$0.97$ on repositories; these are empirical ratios (guarantees in Appendix~\ref{app:game}).

\paragraph{Robustness.}
\label{sec:robustness}
Under the binary outcome (Table~\ref{tab:binary}) information remains the largest effect wherever it was under utility and computation still vanishes for the strongest models. Retrieval stays near zero and specification large when only cleanly applied repository-level edits are kept, and the repository-level computation and feedback effects do not grow with doubled or quadrupled planning budgets or with full failure output (F4). Realising computation as Qwen3-14B's thinking mode raises $\ame{C}$ from $0.110$ to $0.120$, level with information ($0.116$). Feedback fixation survives permuted option orders: the three fixated models request feedback in $63$--$100\%$ of checkpoints under every order (Appendix~\ref{app:robustness}).

\newpage
\bibliographystyle{plainnat}
\bibliography{ref}


\appendix


\section{Discussion and Conclusion}
\label{sec:discussion}

\paragraph{Why feedback is weak, and what follows for allocation.}
Feedback is \emph{artifact-conditional}: it says that the attempt is wrong, and where, but not what right would be, so it cannot correct a misreading of the task, and what it does say the model can largely reconstruct by planning (the negative $\Gamma_{CF}$ terms). It may matter more in an iterative loop, where it also gates termination. For allocation, the first pick matters most and is state-dependent, and models' requests do not track its value; the oracle-to-own gap ($0.05$--$0.22$ at function level, $0.08$--$0.17$ on repositories; Figure~\ref{fig:selfalloc}) bounds what a learned allocator could recover.


\paragraph{Conclusion.}\label{sec:conclusion}
We introduced BRA, a controlled measurement of what a failing agent lacks: it forks a frozen failed state into every combination of computation, information, and feedback, holds a single final revision fixed, and reads off each resource's Banzhaf value and interactions. Applied to six models and two task families, BRA shows that in one revision the binding resource is a reference-derived specification of what the code must do, persistent at the frontier, while the value of planning falls with capability, feedback's is small, and retrieved code is worth nothing; resources substitute, so the first one supplied matters most. The same surfaces score an agent's own choices: three of four models, fixated on feedback, realise less at function level than a random choice would. BRA thus gives a common yardstick for comparing models on the same failures and for training allocation policies against the per-checkpoint optimum.


\section*{Limitations}
BRA measures one revision step. This is a design choice: inside an agent loop the feedback an agent receives depends on its previous edit, so resources can no longer be assigned at a fixed state, and feedback would also confound its information with the extra attempts and the stopping signal it provides; a single step keeps every bundle well defined and the three resources comparable as inputs to the same revision. Our estimates therefore measure feedback as information for one revision, not its value as a verifier or termination signal over a trajectory, which a matched-budget multi-step comparison~\citep{yang2024sweagent} would be needed to establish. The oracle rungs establish what information \emph{would} do, not what a system can obtain; the checkpoints are failures of one weak generator and one weak agent (the native panel shows the same pattern on a strong agent's failures at small $N$); and the within-family ladder removes the vendor confound only up to 32B.

\section*{Ethics Statement}
\label{app:impact}
This work involves no human subjects, crowd workers, or personal data. All tasks come from public benchmarks used under their licenses (MBPP, HumanEval, EvalPlus, SWE-bench Verified), open-weight models are run locally under their licenses, and the remaining models are accessed through APIs under their terms of service. The possible releasing artifacts are failed patches, short text packets, and outcome logs for these public tasks; because the specification packets are derived from the benchmarks' reference solutions, which are already public, releasing them adds no new leakage, but they should not be used as training data for models later evaluated on the same benchmarks. 
We see no direct path to harm beyond that of more capable coding agents in general.

\section{Related Work}
\label{sec:related}

{\bf Test-time scaling of computation.}
Most inference-time work spends more of one resource and reports accuracy against a scalar budget: longer chains of thought~\citep{wei2022cot}, models trained to reason at length~\citep{openai2024o1,deepseek2025r1,muennighoff2025s1}, sampling with voting or a verifier~\citep{wang2023selfconsistency,cobbe2021verifiers,brown2024monkeys,lightman2024verify}, and compute-optimal allocation~\citep{snell2024scaling,wu2024inferencescaling}; its limits are mapped by imperfect verifiers~\citep{stroebl2024flaws,setlur2025verification}, instance-dependent budget curves~\citep{han2025tale}, and end-to-end ablations of agent strategies~\citep{zhu2025agentscaling}. We hold the final revision's budget fixed and vary what precedes it.

{\bf Feedback, information, and allocation.}
Iterative repair feeds execution results or critiques back to the model~\citep{chen2024selfdebug,shinn2023reflexion,madaan2023selfrefine,gou2024critic,gehring2024rlef}; self-repair gains are modest at matched cost and bottlenecked by feedback quality~\citep{olausson2024selfrepair,zhang2024smallverifiers}, and self-correction is rare without an external signal~\citep{huang2024selfcorrect,kamoi2024selfcorrection}. Information is supplied by retrieval~\citep{lewis2020rag,zhang2023repocoder,zhou2023docprompting,xia2024agentless}; \citet{yuan2026memory} run a factorial over memory strategies within one channel. These studies show each channel \emph{can} help but compare pipelines that differ in state, budget, and resource at once. Agent benchmarks report resolution after a full trajectory~\citep{jimenez2024swebench,yang2024sweagent}, increasingly with cost~\citep{kapoor2024agents}; resource acquisition as a decision problem appears in model routing~\citep{chen2023frugalgpt,ong2024routellm}, tool acquisition under heterogeneous costs~\citep{feng2026camdf}, and escalation tests~\citep{wu2026ecoagent}; failure taxonomies are derived from trajectories~\citep{cemri2025mast,fei2026autoresearch}. None intervenes on the resources available at a fixed state, which is what lets us rank the channels, separate retrieved code from a specification, and tell a sufficient resource from a necessary one.


\section{The resource game}
\label{app:game}

\subsection{Proofs of Proposition~\ref{prop:game} and Shapley values}
Write $v(T)=\sum_{S\subseteq T}m(S)$ (M\"obius inversion), $N=\{C,I,F\}$, and for a resource $X$ let $Y,Z$ be the other two.

\paragraph{(i) $\ame{X}=\beta_X$.} $\mathrm{MV}_X(T)=\sum_{S\subseteq T}m(S\cup\{X\})$. Averaging over the four bundles $T\in\{\emptyset,\{Y\},\{Z\},\{Y,Z\}\}$ with equal weight $1/4$ gives $\beta_X=m(X)+\tfrac12 m(XY)+\tfrac12 m(XZ)+\tfrac14 m(XYZ)$; Eq.~\eqref{eq:ame} averages $q_{1if}-q_{0if}$ over the same four bundles with the same weights, so the two coincide. (With $n=3$ the Banzhaf normalisation $2^{-(n-1)}$ is $1/4$.)

\paragraph{(ii) $\Gamma_{XY}$ is the interaction index.} $\Delta_{XY}v(T)=\sum_{S\subseteq T}m(S\cup\{X,Y\})$, so $\Delta_{XY}v(\emptyset)=m(XY)$ and $\Delta_{XY}v(\{Z\})=m(XY)+m(XYZ)$. The Banzhaf interaction index averages these two with weights $2^{-(n-2)}=1/2$; the Shapley interaction index~\citep{grabisch1999interaction} uses weights $|T|!(n-|T|-2)!/(n-1)!$, which for $n=3$ are $0!1!/2!=1/2$ and $1!0!/2!=1/2$. Both equal $m(XY)+\tfrac12 m(XYZ)$, which is $\Gamma_{XY}$ of Eq.~\eqref{eq:ame} (the average of the $XY$ second difference over the two levels of the third factor).

\paragraph{(iii)} $\Gamma_{CIF}$ is the alternating sum over all eight cells, which is $m(N)$ by definition of the M\"obius transform.

\begin{table*}[t]
\centering\small
\caption{Shapley values $\phi_X$ and Banzhaf values $\beta_X=\ame{X}$ of the three resources, averaged over checkpoints. The Shapley values sum exactly to the full-resource gain; the two values differ by at most $0.006$ and rank the resources identically for every model.}
\label{tab:shapley}
\resizebox{\linewidth}{!}{\begin{tabular}{l l r r r r r r r r}
\toprule
Family & Model & $\phi_C$ & $\phi_I$ & $\phi_F$ & $\beta_C$ & $\beta_I$ & $\beta_F$ & $\sum\phi$ & $q_{111}-q_{000}$\\
\midrule
function level & Qwen3-4B & 0.128 & 0.105 & 0.027 & 0.131 & 0.108 & 0.030 & 0.260 & 0.260\\
 & Qwen3-14B & 0.113 & 0.139 & 0.030 & 0.110 & 0.136 & 0.026 & 0.282 & 0.282\\
 & Qwen3-32B & 0.107 & 0.135 & 0.039 & 0.101 & 0.129 & 0.033 & 0.281 & 0.281\\
 & DeepSeek-V3.2 & 0.055 & 0.088 & 0.014 & 0.053 & 0.086 & 0.012 & 0.157 & 0.157\\
 & GPT-5.6 & 0.011 & 0.065 & 0.020 & 0.010 & 0.064 & 0.019 & 0.096 & 0.096\\
 & Claude Sonnet 4.5 & 0.007 & 0.067 & 0.007 & 0.007 & 0.067 & 0.007 & 0.081 & 0.081\\
\midrule
repository level (spec.) & Qwen3-14B & 0.037 & 0.051 & 0.010 & 0.034 & 0.048 & 0.008 & 0.098 & 0.098\\
 & DeepSeek-V3.2 & 0.032 & 0.081 & 0.009 & 0.034 & 0.083 & 0.011 & 0.121 & 0.121\\
 & GPT-5.6 & 0.007 & 0.101 & 0.007 & 0.009 & 0.103 & 0.009 & 0.115 & 0.115\\
 & Claude Sonnet 4.5 & $-$0.023 & 0.089 & 0.002 & $-$0.024 & 0.088 & 0.001 & 0.068 & 0.068\\
\bottomrule
\end{tabular}
}
\end{table*}

\paragraph{(iv) Efficiency.} The Shapley value satisfies $\phi_X=\sum_{S\ni X}m(S)/|S|$, so $\sum_X\phi_X=\sum_{S\neq\emptyset}m(S)=v(N)-v(\emptyset)$. From (i), $\sum_X\beta_X=\sum_{|S|=1}m(S)+\sum_{|S|=2}m(S)+\tfrac34 m(N)=v(N)-v(\emptyset)-\tfrac14 m(N)$.

\paragraph{(v) Bijection.} By (i)--(iii) the vector 
{\small $(q_{000},\ame{C},\ame{I},\ame{F},\Gamma_{CI},\Gamma_{CF},\Gamma_{IF},\Gamma_{CIF})$} 
is an invertible linear function of the eight M\"obius coefficients: $q_{000}=m(\emptyset)$, $\Gamma_{CIF}=m(N)$, $\Gamma_{XY}=m(XY)+\tfrac12 m(N)$ recovers each pairwise coefficient, and $\ame{X}$ then recovers $m(X)$. Since the M\"obius transform is itself a linear bijection of the eight cell means, so is the fingerprint together with $q_{000}$. This is the Banzhaf interaction transform of \citet{grabisch2000equivalent}, restricted to non-empty subsets.

\subsection{Greedy Guarantees}

\paragraph{Submodularity.} For three resources, $v$ is submodular iff $\Delta_{XY}v(T)\le0$ for all pairs and all $T$, i.e. $m(XY)\le0$ and $m(XY)+m(XYZ)\le0$ for each pair. Since $\Gamma_{XY}$ is the mean of these two quantities, submodularity implies $\Gamma_{XY}\le0$ for all pairs, but not conversely.



For a monotone submodular $v$ and additive costs, the greedy rule that repeatedly adds the affordable element of highest marginal value per cost, combined with the best single element, attains at least $(1-1/e)/2$ of the budgeted optimum~\citep{khuller1999budgeted}; with a partial-enumeration refinement the bound is $1-1/e$~\citep{nemhauser1978,khuller1999budgeted}. We use these results only as motivation: Table~\ref{tab:econ} reports the realised greedy-to-optimal ratio directly.

\begin{table*}[t]
\centering\small
\caption{Costs, non-fungibility, and allocation (repository level uses the specification arm for $I$). \emph{Non-fungibility}: within-checkpoint (fixed-effect) five-fold cross-validated $R^2$ for predicting a checkpoint's eight cell means from a binned function of the cell's total token cost (scalar), from the resource indicators (vector), and from both. \emph{Return per cost}: utility gain of each single-resource cell over the empty cell per $1000$ tokens the resource adds to the cell (input and output of the planning and revision calls; the information cost is the packet's tokens, not the oracle's), and per USD at list prices (``--'': local or unpriced model). \emph{Share of surfaces} satisfying the exact submodularity and monotonicity conditions (no noise tolerance). \emph{Greedy / optimal}: mean ratio of the gain of the greedy-by-return rule to that of the optimal affordable bundle, minimum over budgets of 25--75\% and at 100\% of the total cost of all three resources.}
\label{tab:econ}
\resizebox{\linewidth}{!}{\begin{tabular}{l l rrr rrr rrr rr rr}
\toprule
& & \multicolumn{3}{c}{non-fungibility: CV $R^2$} & \multicolumn{3}{c}{$\Delta U$ per 1k tokens} & \multicolumn{3}{c}{$\Delta U$ per USD} & \multicolumn{2}{c}{share of surfaces} & \multicolumn{2}{c}{greedy / optimal}\\
\cmidrule(lr){3-5}\cmidrule(lr){6-8}\cmidrule(lr){9-11}\cmidrule(lr){12-13}\cmidrule(lr){14-15}
Family & Model & scalar & vector & both & $C$ & $I$ & $F$ & $C$ & $I$ & $F$ & submod. & monot. & 25--75\% & 100\%\\
\midrule
function & Qwen3-4B & 0.100 & 0.141 & 0.137 & 0.15 & 0.54 & 0.28 & -- & -- & -- & 0.25 & 0.33 & 0.90 & 0.80\\
 & Qwen3-14B & 0.103 & 0.153 & 0.145 & 0.13 & 1.18 & 0.43 & -- & -- & -- & 0.27 & 0.32 & 0.96 & 0.88\\
 & Qwen3-32B & 0.111 & 0.158 & 0.156 & 0.11 & 1.25 & 0.48 & 325 & 7668 & 2979 & 0.26 & 0.28 & 0.97 & 0.91\\
 & DeepSeek-V3.2 & 0.043 & 0.095 & 0.087 & 0.04 & 0.80 & 0.17 & 36 & 1216 & 267 & 0.44 & 0.46 & 0.99 & 0.97\\
 & GPT-5.6 & 0.038 & 0.040 & 0.051 & 0.03 & 0.60 & 0.19 & -- & -- & -- & 0.62 & 0.61 & 0.99 & 0.99\\
 & Claude Sonnet 4.5 & 0.007 & 0.050 & 0.046 & 0.01 & 0.56 & 0.04 & 2 & 173 & 11 & 0.66 & 0.64 & 0.99 & 0.97\\
\midrule
repository (spec.) & Qwen3-14B & $-$0.001 & 0.023 & 0.005 & 0.01 & 0.11 & 0.18 & -- & -- & -- & 0.78 & 0.81 & 0.95 & 0.94\\
 & DeepSeek-V3.2 & $-$0.010 & 0.036 & 0.024 & 0.01 & 0.15 & 0.03 & 9 & 128 & 29 & 0.53 & 0.57 & 0.93 & 0.89\\
 & GPT-5.6 & $-$0.015 & 0.040 & 0.020 & 0.00 & 0.31 & 0.07 & -- & -- & -- & 0.63 & 0.64 & 0.96 & 0.95\\
 & Claude Sonnet 4.5 & $-$0.006 & 0.036 & 0.026 & 0.00 & 0.23 & 0.08 & 0 & 27 & 13 & 0.60 & 0.52 & 0.97 & 0.93\\
\bottomrule
\end{tabular}
}
\end{table*}

\section{Setup details}
\label{app:setup}

\paragraph{Checkpoint construction.}
Function-level checkpoints are the first attempts of Qwen3-4B in a fixed ``quick draft'' style (one call, $2048$ tokens, temperature $0.7$, seed fixed) on every MBPP+ and HumanEval+ task, kept when the artifact fails the held-out tests. Task IDs, artifacts, dataset versions and hashes (MBPP+ v0.2.0, HumanEval+ v0.1.10), and the selection rule are frozen in a manifest whose hash every outcome row records. Repository-level checkpoints are the submitted patches of a text-based mini-SWE-agent (Qwen3-14B, 75 steps, temperature 0) on a stratified 250-instance frame of SWE-bench Verified, kept when the official harness reports the instance unresolved and the patch is not degenerate (applies, under 200~KB, under 500 changed lines). For each checkpoint we materialise the pre- and post-patch contents of the touched files and the repository's Python sources at the base commit from the benchmark's Docker image.


\begin{table*}[!t]
\centering\small
\caption{Function-level fixed-state panel ($N=194$; held-out utility $U$): empty and full-resource cells, average main effects, each effect's share of their sum, and heterogeneity (share of checkpoints whose best single resource differs from the population's).}
\label{tab:function_main}
\resizebox{\linewidth}{!}{\begin{tabular}{l r r r r r c r}
\toprule
Model & $q_{000}$ & $q_{111}$ & $\mathrm{AME}_C$ & $\mathrm{AME}_I$ & $\mathrm{AME}_F$ & share $C$/$I$/$F$ (\%) & Het.\\
\midrule
Qwen3-4B & 0.338 & 0.597 & \textbf{0.131} {\scriptsize[0.097, 0.164]} & \textbf{0.108} {\scriptsize[0.080, 0.139]} & \textbf{0.030} {\scriptsize[0.011, 0.051]} & 49 / 40 / 11 & 0.30\\
Qwen3-14B & 0.448 & 0.730 & \textbf{0.110} {\scriptsize[0.076, 0.143]} & \textbf{0.136} {\scriptsize[0.105, 0.170]} & \textbf{0.026} {\scriptsize[0.009, 0.043]} & 40 / 50 / 10 & 0.45\\
Qwen3-32B & 0.535 & 0.816 & \textbf{0.101} {\scriptsize[0.074, 0.131]} & \textbf{0.129} {\scriptsize[0.101, 0.159]} & \textbf{0.033} {\scriptsize[0.016, 0.051]} & 39 / 49 / 13 & 0.39\\
DeepSeek-V3.2 & 0.732 & 0.889 & \textbf{0.053} {\scriptsize[0.031, 0.076]} & \textbf{0.086} {\scriptsize[0.059, 0.114]} & \textbf{0.012} {\scriptsize[0.003, 0.021]} & 35 / 57 / 8 & 0.34\\
GPT-5.6 & 0.813 & 0.909 & 0.010 {\scriptsize[$-$0.006, 0.026]} & \textbf{0.064} {\scriptsize[0.038, 0.091]} & \textbf{0.019} {\scriptsize[0.006, 0.033]} & 11 / 69 / 20 & 0.29\\
Claude Sonnet 4.5 & 0.817 & 0.898 & 0.007 {\scriptsize[$-$0.010, 0.025]} & \textbf{0.067} {\scriptsize[0.043, 0.092]} & 0.007 {\scriptsize[$-$0.006, 0.020]} & 8 / 83 / 9 & 0.28\\
\bottomrule
\end{tabular}
}
\end{table*}

\begin{table*}[!t]
\centering\small
\caption{Interactions and conditional necessities, function level. Pairwise interactions are given on all checkpoints and on the $N_{\text{far}}$ checkpoints far from the ceiling ($q_{000}<0.5$ for that model; Holm-adjusted within each sample). Negative pairwise terms indicate substitution. $N_X$ is the loss from removing $X$ from the full bundle.}
\label{tab:function_interactions}
\resizebox{\linewidth}{!}{\begin{tabular}{l r rr rr rr r rrr}
\toprule
& & \multicolumn{2}{c}{$\Gamma_{CI}$} & \multicolumn{2}{c}{$\Gamma_{CF}$} & \multicolumn{2}{c}{$\Gamma_{IF}$} & & & & \\
\cmidrule(lr){3-4}\cmidrule(lr){5-6}\cmidrule(lr){7-8}
Model & $N_{\text{far}}$ & all & far & all & far & all & far & $\Gamma_{CIF}$ & $N_C$ & $N_I$ & $N_F$\\
\midrule
Qwen3-4B & 123 & 0.008 & $-$0.017 & \textbf{$-$0.090} & \textbf{$-$0.132} & 0.028 & 0.027 & $-$0.040 & \textbf{0.080} & \textbf{0.117} & $-$0.011\\
Qwen3-14B & 103 & $-$0.039 & \textbf{$-$0.098} & \textbf{$-$0.056} & \textbf{$-$0.102} & $-$0.027 & \textbf{$-$0.047} & 0.044 & \textbf{0.073} & \textbf{0.114} & $-$0.004\\
Qwen3-32B & 88 & \textbf{$-$0.059} & \textbf{$-$0.123} & \textbf{$-$0.045} & \textbf{$-$0.114} & \textbf{$-$0.028} & \textbf{$-$0.062} & \textbf{0.070} & \textbf{0.067} & \textbf{0.103} & 0.014\\
DeepSeek-V3.2 & 53 & \textbf{$-$0.041} & \textbf{$-$0.103} & $-$0.004 & $-$0.009 & \textbf{$-$0.023} & \textbf{$-$0.074} & 0.027 & \textbf{0.037} & \textbf{0.061} & 0.005\\
GPT-5.6 & 35 & $-$0.026 & \textbf{$-$0.129} & $-$0.002 & \textbf{$-$0.077} & $-$0.019 & \textbf{$-$0.159} & 0.013 & $-$0.001 & \textbf{0.044} & 0.012\\
Claude Sonnet 4.5 & 35 & $-$0.035 & \textbf{$-$0.136} & 0.004 & $-$0.051 & $-$0.009 & \textbf{$-$0.075} & 0.001 & $-$0.008 & \textbf{0.045} & 0.005\\
\bottomrule
\end{tabular}
}
\end{table*}

\paragraph{Information packets and audits.}
Function-level specifications are written by Kimi~K2.5 from the task and the reference solution, in $80$--$140$ words, describing input--output behaviour and edge cases and forbidden to state an algorithm; the audit rejects code fences or code-like lines, any reference-solution line of at least 12 characters, and any held-out or public test literal of at least 6 characters not already present in the task; a rejected draft is regenerated once with the reasons. Repository-level specifications are written from the issue and the reference patch in $90$--$200$ words; the audit additionally rejects diff syntax, any changed line of the reference patch (at least 12 characters and four tokens), and any evaluation test identifier, and, for the behaviour rung, any file path or code-looking function name (containing an underscore or camel case) from the reference patch; class names, plain-English method names, and dunder methods are permitted because a behaviour description cannot avoid the public objects it concerns. The location rung is generated mechanically from the reference patch's hunks and the enclosing definitions in the base-commit sources. All $126$ packets of each rung pass the audit; mean lengths are $155$ (specification), $139$ (behaviour), and $64$ (location) words. Retrieval packets are the top five BM25 chunks (top-level functions, classes, and methods of the repository's Python files at the base commit) for the issue text, excluding regions already shown in the prompt, at most $12{,}000$ characters.


\paragraph{Feedback reports.}
Function-level reports execute the artifact against the dataset's original assertions in a sandbox and report, by level, pass counts (F1), up to five failing inputs (F2), with expected outputs (F3), and with observed outputs (F4). Repository-level reports are derived from the official harness's evaluation of the frozen artifact: outcome counts of failing target and regressed tests, and, under the split-test protocol, the designated observed test's status and a bounded excerpt of its failure output; when the artifact breaks the test run, the report says so and quotes the error lines.

\paragraph{Evaluation.}
Function-level artifacts are scored by EvalPlus in a sandboxed worker on one machine, because per-test time limits are hardware dependent; every condition of an experiment is scored on the same machine. Repository-level artifacts are scored by the SWE-bench harness (version 5.0.2) in the benchmark's own images; a revision that makes the test suite hang is terminated after 1800\,s and scored as a failure.


\section{Repository-level edit protocol}
\label{app:swe}

The revision returns up to six search/replace blocks, each naming a file and quoting at most 25 lines to replace. Blocks are applied to the post-patch files with a fixed fallback order: exact match; match ignoring trailing whitespace; a unique high-similarity window; match ignoring indentation with the replacement re-indented; and, for blocks written as unified-diff hunks, recovery of the old text from the hunk. A block whose search text is not unique is applied at the first occurrence after the previous edit. Partially applicable responses are applied and scored, and the number of discarded blocks is recorded; a response with no applicable block is scored $0$. A response whose applied edits return the files to the base commit is scored deterministically as the base commit (all target tests fail, no regression), since the harness would otherwise evaluate the empty patch. Edits may also target repository files outside the touched set by path. Per row we record the number of blocks, applied, failed, fuzzy-matched, diff-styled, and reverted, so that any sensitivity analysis over the interface can be reproduced.

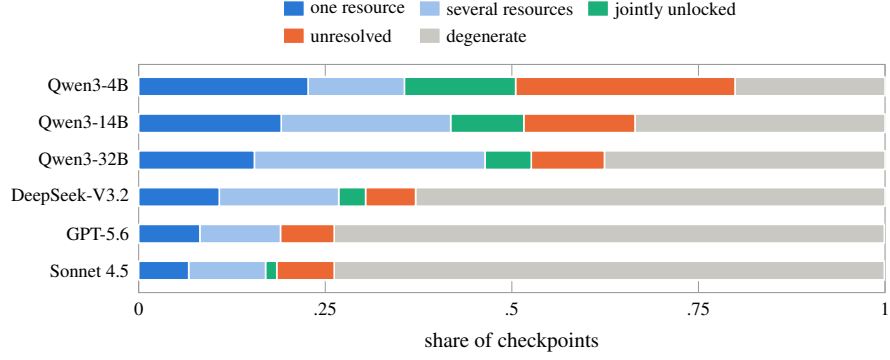
\begin{figure}[t]
\centering
\definecolor{cC}{HTML}{2A78D6}
\definecolor{cI}{HTML}{1BAF7A}
\definecolor{cF}{HTML}{EB6834}
\definecolor{cN}{HTML}{8A8984}
\definecolor{cG}{HTML}{C9C8C3}
\begin{tikzpicture}
\begin{axis}[
  xbar stacked, bar width=7pt, width=0.82\linewidth, height=46mm, every axis/.append style={font=\footnotesize},
  xmin=0, xmax=1, xtick={0,0.25,0.5,0.75,1}, xticklabels={0,.25,.5,.75,1}, xlabel={share of checkpoints},
  ytick={0,...,5}, yticklabels={Qwen3-4B,Qwen3-14B,Qwen3-32B,DeepSeek-V3.2,GPT-5.6,Sonnet 4.5}, y dir=reverse, enlarge y limits=0.12,
  tick label style={font=\scriptsize}, axis line style={black!40}, tick style={black!40},
  legend style={at={(0.5,1.03)}, anchor=south, legend columns=3, draw=none, font=\scriptsize, /tikz/every even column/.append style={column sep=3pt}},
  legend cell align=left, reverse legend=false,
]
\addplot[fill=cC, draw=white, line width=0.6pt] coordinates {(0.227,0) (0.191,1) (0.155,2) (0.108,3) (0.082,4) (0.067,5)};
\addlegendentry{one resource}
\addplot[fill=cC!45, draw=white, line width=0.6pt] coordinates {(0.129,0) (0.227,1) (0.309,2) (0.160,3) (0.108,4) (0.103,5)};
\addlegendentry{several resources}
\addplot[fill=cI, draw=white, line width=0.6pt] coordinates {(0.149,0) (0.098,1) (0.062,2) (0.036,3) (0.000,4) (0.015,5)};
\addlegendentry{jointly unlocked}
\addplot[fill=cF, draw=white, line width=0.6pt] coordinates {(0.294,0) (0.149,1) (0.098,2) (0.067,3) (0.072,4) (0.077,5)};
\addlegendentry{unresolved}
\addplot[fill=cG, draw=white, line width=0.6pt] coordinates {(0.201,0) (0.335,1) (0.376,2) (0.629,3) (0.737,4) (0.737,5)};
\addlegendentry{degenerate}
\end{axis}
\end{tikzpicture}
\vspace{-2mm}
\caption{Label prevalence, function level ($\tau=0.2$, ceiling $0.8$). ``One resource'' and ``several resources'' are checkpoints responsive to exactly one or to more than one single resource.}
\label{fig:labels}
\end{figure}

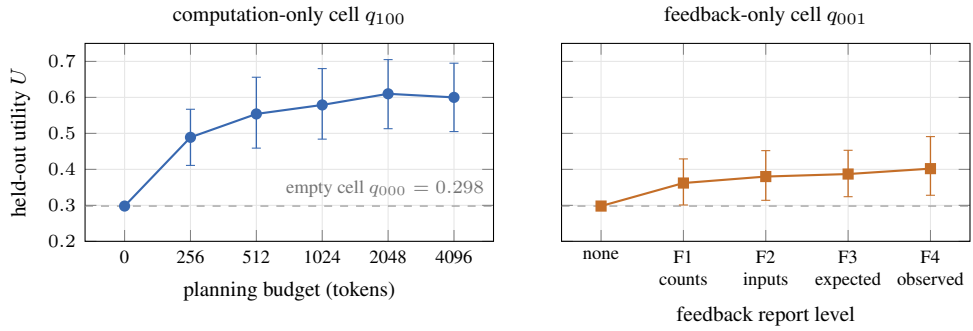
\begin{figure*}[t]
\centering
\definecolor{cC}{RGB}{56,104,176}
\definecolor{cF}{RGB}{196,110,40}
\begin{tikzpicture}
\begin{groupplot}[
  group style={group size=2 by 1, horizontal sep=9mm, ylabels at=edge left, yticklabels at=edge left},
  width=0.5\linewidth, height=42mm, every axis/.append style={font=\footnotesize},
  ymin=0.2, ymax=0.75, ytick={0.2,0.3,0.4,0.5,0.6,0.7},
  ylabel={held-out utility $U$}, ylabel near ticks,
  xtick=data, enlarge x limits=0.12,
  grid=major, grid style={black!10},
  tick label style={font=\scriptsize}, xticklabel style={align=center},
  title style={font=\footnotesize, yshift=-1mm},
  every axis plot/.append style={line width=0.8pt, mark size=1.7pt},
  error bars/y dir=both, error bars/y explicit, error bars/error bar style={line width=0.5pt},
  clip=false,
]
\nextgroupplot[title={computation-only cell $q_{100}$}, xlabel={planning budget (tokens)},
  symbolic x coords={0,256,512,1024,2048,4096}]
\draw[dashed, black!45] ({rel axis cs:0,0}|-{axis cs:0,0.298}) -- ({rel axis cs:1,0}|-{axis cs:0,0.298});
\node[anchor=south east, font=\scriptsize, text=black!55] at ({rel axis cs:1,0}|-{axis cs:0,0.298}) {empty cell $q_{000}=0.298$};
\addplot[color=cC, mark=*] table[x=x, y=y, y error plus=hi, y error minus=lo] {
    x y lo hi
    0 0.298 0.000 0.000
    256 0.489 0.078 0.078
    512 0.554 0.095 0.102
    1024 0.579 0.095 0.101
    2048 0.610 0.097 0.095
    4096 0.600 0.095 0.095
};
\nextgroupplot[title={feedback-only cell $q_{001}$}, xlabel={feedback report level},
  symbolic x coords={none,F1,F2,F3,F4},
  xticklabels={none, {F1\\counts}, {F2\\inputs}, {F3\\expected}, {F4\\observed}}]
\draw[dashed, black!45] ({rel axis cs:0,0}|-{axis cs:none,0.298}) -- ({rel axis cs:1,0}|-{axis cs:none,0.298});
\addplot[color=cF, mark=square*] table[x=x, y=y, y error plus=hi, y error minus=lo] {
    x y lo hi
    none 0.298 0.000 0.000
    F1 0.362 0.061 0.067
    F2 0.380 0.066 0.072
    F3 0.387 0.063 0.066
    F4 0.402 0.074 0.089
};
\end{groupplot}
\end{tikzpicture}
\vspace{-2mm}
\caption{Dose curves on the 40 hardest function-level checkpoints (DeepSeek-V3.2, $k=5$; empty cell $q_{000}=0.298$, dashed). Left: the computation-only cell $q_{100}$ against the planning budget. Right: the feedback-only cell $q_{001}$ against the feedback report level. Error bars are 95\% checkpoint-bootstrap intervals of the paired gain over the empty cell, drawn around the mean.}
\label{fig:dose}
\end{figure*}

\section{Deviations from the pre-registered protocol}
\label{app:deviations}

Every change after freezing is logged in an append-only ledger with a timestamp and reason; the entries relevant to the reported results are summarised here. (1)~The planning budget was raised from $1024$ to $2048$ tokens after a shakedown showed $20\%$ of plans truncated at $1024$; at $2048$, $5\%$ of function-level and $22\%$ of repository-level plans are truncated. (2)~Code extraction from revisions was made robust to prose around fenced code after one model returned explanations with its code; the fix is applied retroactively to cached outputs of all models. (3)~The repository-level edit matcher gained the fallbacks of Appendix~\ref{app:swe} after a shakedown; all reported repository-level runs use the final protocol, and the two earlier partial launches are void. (4)~Harness time-outs caused by a revision were initially treated as evaluator errors and are now scored as failures; two runs were resumed under the corrected rule with cached model calls. (5)~The behaviour-only audit was relaxed to permit class names, plain-English method names, and dunder methods after it rejected packets that could not describe the behaviour otherwise; the relaxation is stated in Appendix~\ref{app:setup}. (6)~Qwen3-32B is served through a cloud API rather than locally because the local accelerators were occupied; sampling parameters are identical to the local tiers. (7)~The information-dose ladder and the repository-level oracle arm were added after the main runs, in response to the null retrieval result, and are labelled as additions; the Qwen3 ladder, the native-failure panel, and the self-allocation probe were planned for a second stage and run after the main runs. (8)~Before the repository-level repair runs, the checkpoint generator was changed from DeepSeek-V3.2 to Qwen3-14B, because DeepSeek-V3.2 resolved most instances within 75 steps and left too few unresolved patches; its 30 unresolved patches form the native-failure panel.


\begin{table*}[!t]
\centering\small
\caption{Repository-level robustness. (a) Interactions under the retrieval and specification arms ($N=126$). (b) Main effects by artifact stratum; ``retr.'' and ``spec.'' are the information effect under the two arms. (c) Native-failure panel.}
\label{tab:swe_robust}
\resizebox{0.9\linewidth}{!}{\begin{tabular}{@{}l@{}}
\textbf{(a) Interactions} ($N=126$)\\[2pt]
\begin{tabular}{l rrrr rrrr}
\toprule
& \multicolumn{4}{c}{retrieval information} & \multicolumn{4}{c}{audited specification}\\
\cmidrule(lr){2-5}\cmidrule(lr){6-9}
Model & $\Gamma_{CI}$ & $\Gamma_{CF}$ & $\Gamma_{IF}$ & $\Gamma_{CIF}$ & $\Gamma_{CI}$ & $\Gamma_{CF}$ & $\Gamma_{IF}$ & $\Gamma_{CIF}$\\
\midrule
Qwen3-14B & $-$0.021 & $-$0.024 & $-$0.005 & 0.027 & 0.015 & $-$0.021 & 0.009 & 0.033\\
DeepSeek-V3.2 & $-$0.006 & $-$0.012 & $-$0.018 & $-$0.010 & $-$0.015 & $-$0.022 & 0.017 & $-$0.030\\
GPT-5.6 & 0.028 & $-$0.021 & $-$0.008 & $-$0.002 & $-$0.001 & $-$0.032 & 0.018 & $-$0.023\\
Claude Sonnet 4.5 & \textbf{0.053} & $-$0.040 & 0.009 & $-$0.031 & $-$0.024 & $-$0.019 & $-$0.007 & 0.011\\
\bottomrule
\end{tabular}\\[6pt]
\rule{0pt}{14pt}\textbf{(b) Artifact strata}: the agent's patch breaks the test run ($N=57$) or target tests fail ($N=69$)\\[2pt]
\begin{tabular}{l l r r r r r r}
\toprule
Model & stratum & $q_{000}$ & $q_{111}$ & $\mathrm{AME}_C$ & $\mathrm{AME}_I$ (retr.) & $\mathrm{AME}_I$ (spec.) & $\mathrm{AME}_F$\\
\midrule
Qwen3-14B & test run broken & 0.041 & 0.023 & $-$0.009 & 0.000 & 0.019 & $-$0.000\\
 & tests execute & 0.014 & 0.088 & \textbf{0.036} & 0.017 & \textbf{0.073} & 0.001\\
DeepSeek-V3.2 & test run broken & 0.221 & 0.257 & 0.043 & $-$0.004 & \textbf{0.104} & $-$0.005\\
 & tests execute & 0.252 & 0.289 & 0.035 & 0.016 & \textbf{0.066} & $-$0.007\\
GPT-5.6 & test run broken & 0.468 & 0.412 & 0.004 & $-$0.029 & \textbf{0.096} & $-$0.026\\
 & tests execute & 0.385 & 0.475 & \textbf{0.040} & \textbf{0.033} & \textbf{0.108} & 0.014\\
Claude Sonnet 4.5 & test run broken & 0.294 & 0.310 & 0.026 & 0.003 & \textbf{0.097} & $-$0.006\\
 & tests execute & 0.349 & 0.360 & 0.005 & $-$0.006 & \textbf{0.081} & 0.021\\
\bottomrule
\end{tabular}\\[6pt]
\rule{0pt}{14pt}\textbf{(c) Native failures}: DeepSeek-V3.2 repairs 30 checkpoints on which it had itself failed as the agent ($k=3$)\\[2pt]
\begin{tabular}{l r r r r r}
\toprule
Information arm & $q_{000}$ & $q_{111}$ & $\mathrm{AME}_C$ & $\mathrm{AME}_I$ & $\mathrm{AME}_F$\\
\midrule
retrieval & 0.097 & 0.115 & 0.030 {\scriptsize[$-$0.024, 0.104]} & $-$0.004 {\scriptsize[$-$0.046, 0.029]} & $-$0.007 {\scriptsize[$-$0.042, 0.021]}\\
audited specification & 0.097 & 0.187 & \textbf{0.047} {\scriptsize[0.002, 0.099]} & \textbf{0.079} {\scriptsize[0.013, 0.159]} & $-$0.027 {\scriptsize[$-$0.063, 0.003]}\\
\bottomrule
\end{tabular}
\end{tabular}
}
\end{table*}

\section{Compute}
\label{app:cost}



Qwen3-4B and Qwen3-14B are served locally with vLLM on one node with three NVIDIA A100 40\,GB GPUs, 96 CPU cores, and 1.1\,TB of memory; all other models are called through APIs. Evaluation runs on the same node, EvalPlus in sandboxed CPU workers and the SWE-bench harness in the benchmark's Docker images. With concurrent calls, a function-level fixed-state panel (194 checkpoints, eight cells, $k$ rollouts) completes in one to two hours of wall-clock time and a repository-level arm (126 checkpoints, $k=3$) in two to nine hours, dominated by harness evaluation.

\section{Additional results}
\label{app:tables}
Bold entries follow the rule of \S\ref{sec:experiments}: two-sided checkpoint-bootstrap $p<0.05$, Holm-adjusted within model for the three pairwise interactions, unadjusted otherwise. Bracketed ranges are 95\% intervals.

\subsection{Function level}
Table~\ref{tab:function_main} gives the main effects behind Figure~\ref{fig:resource_effects}, Table~\ref{tab:function_interactions} the interactions and conditional necessities, Figure~\ref{fig:ceiling} the two ceiling checks of \S\ref{sec:q2}--\ref{sec:q3}, Figure~\ref{fig:labels} the label prevalence, Figure~\ref{fig:dose} the dose curves, Table~\ref{tab:placebo} the placebo, and Table~\ref{tab:function_cells} all eight cell means.

\noindent\textbf{Cells and conditional necessities.} Even the full bundle leaves much unsolved ($q_{111}=0.60$ for Qwen3-4B, at most $0.91$ for any model; Table~\ref{tab:function_cells}). Information alone is the best single cell for every model but Qwen3-4B, for which computation alone is ($0.499$ against $0.418$), and for GPT-5.6 and Sonnet~4.5 the information-only cell is within $0.01$ of the full bundle ($0.903$ against $0.909$; $0.906$ against $0.898$): once information is present, the other two resources add almost nothing. Removing feedback from the full bundle has no significant cost for any model ($N_F$ between $-0.011$ and $0.014$), removing computation matters only for the four weaker models, and removing information matters for all (Table~\ref{tab:function_interactions}). The three-way term is significant only for Qwen3-32B.

\begin{table}[t]
\centering\small
\caption{Placebo check, function level (paired by checkpoint). The placebo is a length-matched neutral text supplied in place of the specification.}
\label{tab:placebo}
\resizebox{0.7\linewidth}{!}{\begin{tabular}{l rr}
\toprule
Contrast & Qwen3-14B & DeepSeek-V3.2\\
\midrule
placebo $-$ nothing & $-$0.016 {\scriptsize[$-$0.040, 0.007]} & $-$0.002 {\scriptsize[$-$0.020, 0.016]}\\
specification $-$ placebo & \textbf{0.195} {\scriptsize[0.149, 0.245]} & \textbf{0.126} {\scriptsize[0.089, 0.164]}\\
specification $-$ nothing & \textbf{0.180} {\scriptsize[0.134, 0.228]} & \textbf{0.124} {\scriptsize[0.086, 0.162]}\\
\bottomrule
\end{tabular}
}
\end{table}

\begin{table}[t]
\centering\small
\caption{All eight cell means, function level.}
\label{tab:function_cells}
\resizebox{0.8\linewidth}{!}{\begin{tabular}{l rrrrrrrr}
\toprule
Model & 000 & 001 & 010 & 011 & 100 & 101 & 110 & 111\\
\midrule
Qwen3-4B & 0.338 & 0.389 & 0.418 & 0.517 & 0.499 & 0.481 & 0.608 & 0.597\\
Qwen3-14B & 0.448 & 0.526 & 0.627 & 0.657 & 0.616 & 0.616 & 0.734 & 0.730\\
Qwen3-32B & 0.535 & 0.622 & 0.725 & 0.749 & 0.706 & 0.713 & 0.801 & 0.816\\
DeepSeek-V3.2 & 0.732 & 0.764 & 0.856 & 0.851 & 0.814 & 0.828 & 0.884 & 0.889\\
GPT-5.6 & 0.813 & 0.846 & 0.903 & 0.910 & 0.841 & 0.865 & 0.898 & 0.909\\
Claude Sonnet 4.5 & 0.817 & 0.827 & 0.906 & 0.906 & 0.840 & 0.853 & 0.893 & 0.898\\
\bottomrule
\end{tabular}
}
\end{table}

\noindent\textbf{Ceiling checks.} On the matched subset (Figure~\ref{fig:ceiling}, top) the information effect grows for every model, to $0.20$--$0.25$ for all but Qwen3-4B ($0.117$), while computation shrinks for the three Qwen models (e.g.\ $0.110\to0.056$ for Qwen3-14B). Away from the ceiling (bottom) substitution is not specific to computation: $\Gamma_{IF}$ also becomes significantly negative for five of six models. Both subsets are selected on the empty cell, which also enters every contrast; checkpoints whose empty cell is low by chance are over-represented, which inflates all three main effects equally and pushes the interactions down. These panels therefore test whether the ceiling explains the pattern, not the exact size of the effects.

\noindent\textbf{Labels.} The share of degenerate checkpoints ($q_{000}\ge0.8$) rises from $0.20$ for Qwen3-4B to $0.74$ for GPT-5.6 and Sonnet~4.5 (Figure~\ref{fig:labels}), which is the ceiling the matched subset removes. Among checkpoints that a single resource unlocks, computation dominates for Qwen3-4B ($0.19$ of $0.23$) and information for GPT-5.6 and Sonnet ($0.06$ of $0.08$ and of $0.07$); feedback alone unlocks at most $1.5\%$ of checkpoints for any model. Jointly unlocked checkpoints fall from $0.15$ for Qwen3-4B to at most $0.04$ for the three strongest models.

\noindent\textbf{Doses and placebo.} On the dose subset computation saturates early (Figure~\ref{fig:dose}): $256$ tokens already give $61\%$ of the $2048$-token gain and $512$ tokens $82\%$, and $4096$ tokens add nothing ($0.600$ against $0.610$). Feedback gains most from its first level (pass counts, $+0.064$) and at most $0.02$ from each further level. In Table~\ref{tab:placebo} the specification-minus-nothing contrast ($0.180$, $0.124$) is the simple effect at $C{=}F{=}0$; it exceeds the average effect ($\ame{I}=0.136$, $0.086$) because the average also covers bundles in which computation or feedback has already recovered part of the same failures.





\subsection{Repository level}
Table~\ref{tab:swe_robust} complements Table~\ref{tab:swe_main} with interactions, artifact strata, and the native-failure panel; the information dose ladder is Figure~\ref{fig:information_types}.

\noindent\textbf{Repository-level robustness.} Under the specification arm no interaction is significant and none exceeds $0.033$ in absolute value (Table~\ref{tab:swe_robust}a). By stratum (b), the specification effect is positive for every model and similar in the two strata for DeepSeek-V3.2, GPT-5.6, and Sonnet~4.5 ($0.066$--$0.108$).


\subsection{Robustness}
\label{app:robustness}

\noindent\textbf{Outcome definition.} Under the binary ``all held-out tests pass'' outcome (Table~\ref{tab:binary}) information remains the largest effect for the same five function-level models and every repository-level specification arm; at function level computation vanishes for GPT-5.6 and Sonnet~4.5. At function level the information effect, and the computation effect of the weaker models, are larger than under utility (e.g.\ $\ame{C}=0.212$ against $0.131$ for Qwen3-4B).

\noindent\textbf{Thinking mode.} Realising computation as Qwen3-14B's thinking mode on the revision call (at most $8192$ tokens, temperature $0.6$) instead of a planning call, on the same 194 checkpoints with $k=3$, gives $\ame{C}=0.120$ [0.091, 0.150] against $\ame{C}=0.110$ for the planning call, with $\ame{I}=0.116$ and $\ame{F}=0.047$ (planning call: $0.136$, $0.026$) and $\Gamma_{CI}=-0.066$. Thinking exhausted the budget and returned no code in 95 of 2328 thinking cells; under the fixed-budget rule these are scored $0$.

\begin{table}[t]
\centering\small
\caption{Main effects under the binary outcome (all held-out tests pass). Information remains the largest effect wherever it was under utility; at function level computation vanishes for GPT-5.6 and Sonnet~4.5.}
\label{tab:binary}
\resizebox{0.75\linewidth}{!}{\begin{tabular}{l l r r r}
\toprule
Panel & Model & $\mathrm{AME}_C$ & $\mathrm{AME}_I$ & $\mathrm{AME}_F$\\
\midrule
function & Qwen3-4B & \textbf{0.212} & \textbf{0.155} & \textbf{0.032}\\
 & Qwen3-14B & \textbf{0.174} & \textbf{0.200} & \textbf{0.024}\\
 & Qwen3-32B & \textbf{0.162} & \textbf{0.192} & 0.022\\
 & DeepSeek-V3.2 & \textbf{0.067} & \textbf{0.158} & 0.009\\
 & GPT-5.6 & 0.010 & \textbf{0.136} & 0.009\\
 & Claude Sonnet 4.5 & 0.003 & \textbf{0.140} & 0.007\\
\midrule
repo., retrieval & Qwen3-14B & 0.016 & 0.009 & $-$0.000\\
 & DeepSeek-V3.2 & 0.038 & 0.003 & $-$0.005\\
 & GPT-5.6 & 0.024 & 0.003 & $-$0.001\\
 & Claude Sonnet 4.5 & 0.007 & $-$0.002 & 0.013\\
\midrule
repo., specification & Qwen3-14B & \textbf{0.033} & \textbf{0.048} & 0.007\\
 & DeepSeek-V3.2 & \textbf{0.034} & \textbf{0.083} & 0.011\\
 & GPT-5.6 & 0.010 & \textbf{0.105} & 0.010\\
 & Claude Sonnet 4.5 & \textbf{$-$0.028} & \textbf{0.089} & 0.003\\
\bottomrule
\end{tabular}
}
\end{table}

\noindent\textbf{Computation dose on repositories.} $22\%$ of repository-level plans hit the $2048$-token budget. Re-running the computation-only cell with $4096$ and $8192$ tokens (paired over 126 checkpoints, $k=3$) changes it by $+0.014$ [$-0.022$, $+0.052$] and $+0.002$ [$-0.030$, $+0.034$] for DeepSeek-V3.2 and by $+0.007$ [$-0.026$, $+0.040$] and $-0.023$ [$-0.055$, $+0.006$] for GPT-5.6. Neither model gains from planning beyond $2048$ tokens, consistent with the saturation at function level.

\noindent\textbf{Feedback dose on repositories.} On the 33 split-test instances, replacing the F3 excerpt by the full failure output of the observed test and the names of regressed tests (F4) changes the feedback-only cell by $-0.008$ [$-0.066$, $+0.051$] and the retrieval-plus-feedback cell by $-0.043$ [$-0.104$, $+0.013$] for DeepSeek-V3.2; more detailed feedback brings no detectable gain on this subset.

\noindent\textbf{Edit interface.} Repository-level revisions are search/replace edits; cells with no applicable edit are scored $0$, and partial applications and reverts to the base commit are scored as applied; all three are flagged. Restricting to checkpoints on which no revision in any of the eight cells had a failed edit block, an empty edit, or a revert (GPT-5.6: 71 retrieval and 65 specification checkpoints; Sonnet~4.5: 66 and 53) leaves retrieval at $0.005$ and $-0.029$ and raises the specification effect to $0.131$ and $0.145$; for DeepSeek-V3.2 and Qwen3-14B too few checkpoints remain (at most 31) for a reliable estimate. Since reverts reflect the model's judgement, this is a sensitivity check, not an estimate (Appendix~\ref{app:swe}).

\noindent\textbf{Option order.} Repeating the probe with the options presented as (feedback, none, compute, information) and as (information, compute, none, feedback): the three fixated models request feedback in $92$--$99\%$ (Sonnet), $63$--$99\%$ (DeepSeek), and $87$--$100\%$ (Qwen3-14B) of checkpoints across the three orders and both task families; GPT-5.6 requests information $5$--$49\%$ and feedback $30$--$93\%$. Feedback remains the majority choice of the three fixated models under every order.

\end{document}